\documentclass[twocolumn]{autart}
\usepackage{enumitem}

\usepackage{balance}
\usepackage{url} 
\usepackage{booktabs} 
\usepackage[utf8]{inputenc}
\usepackage{amsmath}
\usepackage{amsfonts}
\usepackage{amssymb}
\usepackage{theorem}
\usepackage{graphicx}

\usepackage{subcaption}
\usepackage{color}
\usepackage{cite}

\newcommand{\revone}[1]{{\color{black}#1}}
\newcommand{\revtwo}[1]{{\color{black}#1}}
\newcommand{\revtwoarxiv}[1]{{\color{black}#1}}

{\theorembodyfont{\upshape}\newtheorem{theorem}{Theorem}}
{\theorembodyfont{\upshape}\newtheorem{proposition}{Proposition}}
{\theorembodyfont{\upshape}\newtheorem{lemma}{Lemma}}
{\theorembodyfont{\upshape}\newtheorem{corollary}{Corollary}}
{\theorembodyfont{\upshape}\newtheorem{definition}{Definition}}
{\theorembodyfont{\itshape}}
{\theorembodyfont{\upshape}\newtheorem{remark}{Remark}}
{\theorembodyfont{\upshape}}
{\theorembodyfont{\upshape}}
{\theorembodyfont{\upshape}\newtheorem{assumption}{Assumption}}

\graphicspath{ {figs/} }

\begin{document}

\begin{frontmatter}
\title{Modeling and Passivity Properties of Multi-Producer District Heating Systems\thanksref{footnoteinfo}}

\thanks[footnoteinfo]{
Corresponding author J.~E.~Machado.}

\author[Groningen]{Juan E. Machado}\ead{j.e.machado.martinez@rug.nl},    
\author[Pavia,Groningen]{Michele Cucuzzella}\ead{michele.cucuzzella@unipv.it},               
\author[Groningen]{Jacquelien M. A. Scherpen}\ead{j.m.a.scherpen@rug.nl}  

\address[Groningen]{
Faculty of Science and Engineering, University of Groningen, the
Netherlands}  
\address[Pavia]{Department of Electrical, Computer and Biomedical Engineering, University of Pavia, 
Italy}


\begin{abstract}                          
We propose a comprehensive nonlinear ODE-based thermo-hydraulic model of a district heating system featuring several heat producers, consumers and storage devices which are interconnected through a distribution network of meshed topology whose temperature dynamics are explicitly considered.
Moreover, we analyze the conditions under which the hydraulic and thermal subsystems of the model exhibit shifted passivity properties. 
\revone{For the hydraulic subsystem, our claims on passivity draw on the monotonicity of the vector field associated to the \revtwo{district heating} system's flow dynamics, which mainly codifies viscous friction effects on the system's pressures.}  For the temperature dynamics, we propose a storage function based on the {\em ectropy function} of  a thermodynamic system, recently used in the passivity analysis of heat exchanger networks.
%
%
\end{abstract}

\end{frontmatter}

\section{Introduction}



District heating (DH) has been identified as a key technology to enable  the heating sector's  potential to  reduce  greenhouse emissions  due  to the  possibility to seamlessly	 include environmentally friendly energy sources and storage devices \cite{Lund2014,Werner2017,
vandermeulen_control_review_18,
dominkovic_cluster_20}.
A DH system comprises a network of pipes connecting buildings in a neighborhood, town center or whole city, so that they can be served from varied heat production units \cite{Lund2014}. To further unlock the potential of DH systems, prospective installations will feature multiple, distributed heat sources,  {\em e.g.,}  waste-to-energy facilities or solar collectors \cite{Lund2014},   promoting as a consequence heat distribution networks of meshed topology, as opposed to the salient tree-like structure of conventional installations with a single heat source \cite{wang_meshed_17,vesterlund_multisource_2017}.
%
%


\revone{Modeling aspects of DH systems with a single heat producer has received considerable attention. Following a rigorous graph-theoretic approach, a  control system based on ordinary differential equations (ODE) and focused  on the  hydraulic layer, is presented in  \cite{DePersis2011}. Volume and temperature dynamics are modeled in \cite{Scholten_tcst_2015} for a DH system based on a single stratified water storage tank adjacent to the heat production plant.  Comprehensive,  partial differential equation (PDE)-based DH pipe models are presented in  \cite{krug_DH_2021,hauschild_ph_20,rein_rom_2021}: a fully discretized model is subsequently obtained in  \cite{krug_DH_2021} to be used in the design of a nonlinear optimal controller;  semi-discrete and PDE-based DH port-Hamiltonian models \cite{arjan_dimitri_book_PH} are formulated in \cite{hauschild_ph_20}; and model reduction is performed on a semi-discretized DH system model in \cite{rein_rom_2021}.  
}
\revone{
Modeling of DH systems with multiple heat producers has been addressed  in  \cite{wang_meshed_17,
vesterlund_multisource_2017,
Trip2019a,Alisic2019}. (Static) steady-sate hydraulic and thermo-hydraulic  modeling is considered in \cite{wang_meshed_17} and \cite{vesterlund_multisource_2017}, for the purposes of solving operational optimization problems. Dynamic volume storage modeling (and control) is presented in \cite{Trip2019a} for a DH system with multiple storage tanks. A similar model is established in \cite{Alisic2019}, which further considers temperature  dynamics.\footnote{We refer the reader to \cite{talebi_review_16} and \cite{guelpa_storage_review_2019} for surveys on DH  modeling and thermal energy storage in DH applications, respectively.}
}
%
%

\revone{
Passivity analysis within the context of heating networks has been considered in  \cite{mukherjee_building_2012,dong_passivity_19}.  In \cite{mukherjee_building_2012},  a (linear) model to describe the temperature dynamics of a multi-zone building is presented and subsequently shown to be passive via a storage function which is quadratic \revtwo{in} the rooms' temperatures. In \cite{dong_passivity_19}, a general model of a network of heat exchangers  is shown to be {\em shifted} passive  using a novel storage function based on the concept of ectropy  \cite{haddad_book_new}. It was mentioned in a previous paragraph that port-Hamiltonian formulations of (single producer) DH system models are presented in \cite{hauschild_ph_20}, thus, passivity follows directly under mild assumptions \cite{arjan_dimitri_book_PH}.\footnote{\revone{Ectropy  is a quadratic function on the total energy of a thermodynamic system and is described in \cite{haddad_book_new} as the  dual of  entropy in the sense that it represents a measure of the tendency of a thermodynamic system to do useful work and grow more organized  (see also  \cite{willems_review_haddad_06}). On the other hand, shifted passivity is particularly relevant in these applications by allowing the stability assessment and stabilization of \revtwo{non-trivial} equilibria (see \cite{bayu_scl_07,pooya_aut_19,nima_scl_19}).}}
}

%

{\bf Contributions:} \revone{In this paper, we build  on  \cite{DePersis2011,
Scholten_tcst_2015,wang_meshed_17,
hauschild_ph_20,krug_DH_2021,rein_rom_2021} and  present a  simplified ODE-based model} that simultaneously  describes the hydraulic and thermal dynamic behavior  of a DH system featuring multiple heat producers, storage devices and consumers, all of them  connected to a common meshed distribution network.   The model is highly nonlinear  due to  the consideration of frictional \revone{forces} in the flow dynamics and due to the coupling between the flow and temperature dynamics.\footnote{\revone{Our  modeling procedure  is based on  \cite{DePersis2011,valdimarsson_14,Scholten_tcst_2015,wang_meshed_17,
hauschild_ph_20,krug_DH_2021} and every important difference is emphasized throughout our paper (see, {\em e.g.}, Remarks~\ref{rem:diff_claudios_etc} and \ref{rem:thermal_flows_wlog}).}}
As a second contribution we analyze the conditions under which flow and thermal dynamics of the proposed model are shifted passive \revone{and briefly touch on the implications} for the design of decentralized passivity-based controllers with stability guarantees. 
\revone{Our claims on the passivity of the DH system's flow dynamics are based on the observations made in \cite{DePersis2011}, in the single producer setting,  about the monotonicity of the associated vector field.} On the other hand,  following \cite{dong_passivity_19} (see also \cite{hauschild_ph_20}), for the thermal dynamics we propose a quadratic storage function based on the total ectropy,  extending the results of \cite{dong_passivity_19,hauschild_ph_20} to the multiple heat producers and multiple storage devices case.

{\bf Notation:} The symbol $\mathbb{R}$ denotes the set of real numbers. For a vector $x\in\mathbb{R}^n$,  $x_i$ denotes its $i$th component, {\em i.e.}, $x=[x_1,\dots, x_n]^\top$; moreover,  $\mathbf{sign}(x)=[\text{sign}(x_1),\dots,\text{sign}(x_n)]^\top $, with $\text{sign}(0)=0$,  and $\vert x \vert = [\vert x_1 \vert,\dots, \vert x_n \vert]^\top$. An $m\times n$ matrix with all-zero entries is written as $\boldsymbol{0}_{m\times n}$. An $n$-vector of ones is written as $\boldsymbol{1}_n$, whereas the identity matrix of size $n$ is represented by $I_n$. For any vector $x\in \mathbb{R}^n$, we denote by \revone{$\mathrm{diag} (x) $}  a diagonal matrix with elements $x_i$ in its main diagonal. For any time-varying signal $w$, we represent by $\bar w$ its steady-state value, if exists. \revone{Also, we write time derivatives as $\dot{x}(t)$, and omit the argument $t$ whenever is clear from  the context.}

\section{System Setup}\label{sec:system_model}

\begin{figure}
\begin{center}
\includegraphics[width=0.8\linewidth]{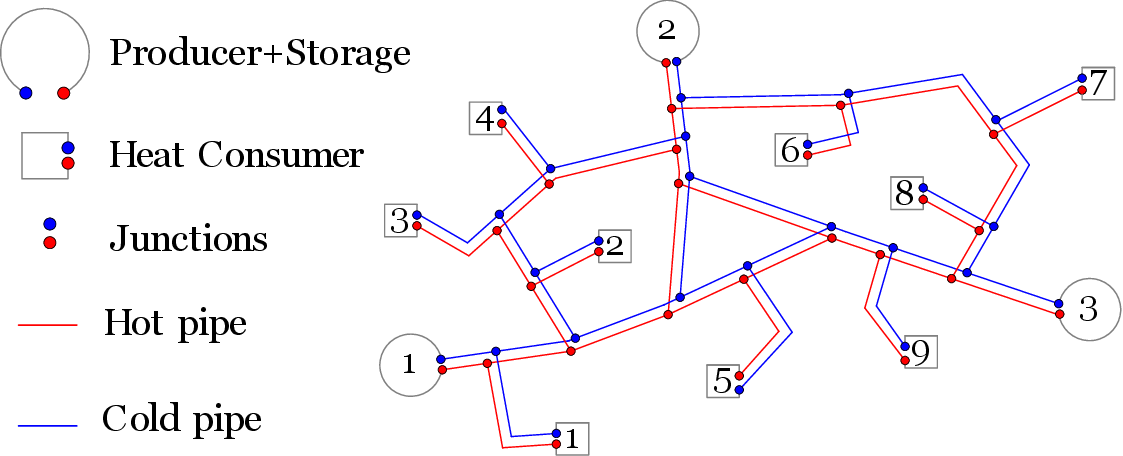}
\caption{\revone{Sketch of a DH system in which a distribution network interconnects 3 heat producers and 9 consumers \cite{wang_meshed_17}. Each producer is interfaced to the distribution network through a storage tank (see Fig.~\ref{fig:prod_stor} for details).}}
\label{fig:dh_meshed}
\end{center}
\end{figure}

\begin{figure}[t]
\begin{center}
\includegraphics[width=0.8\linewidth]{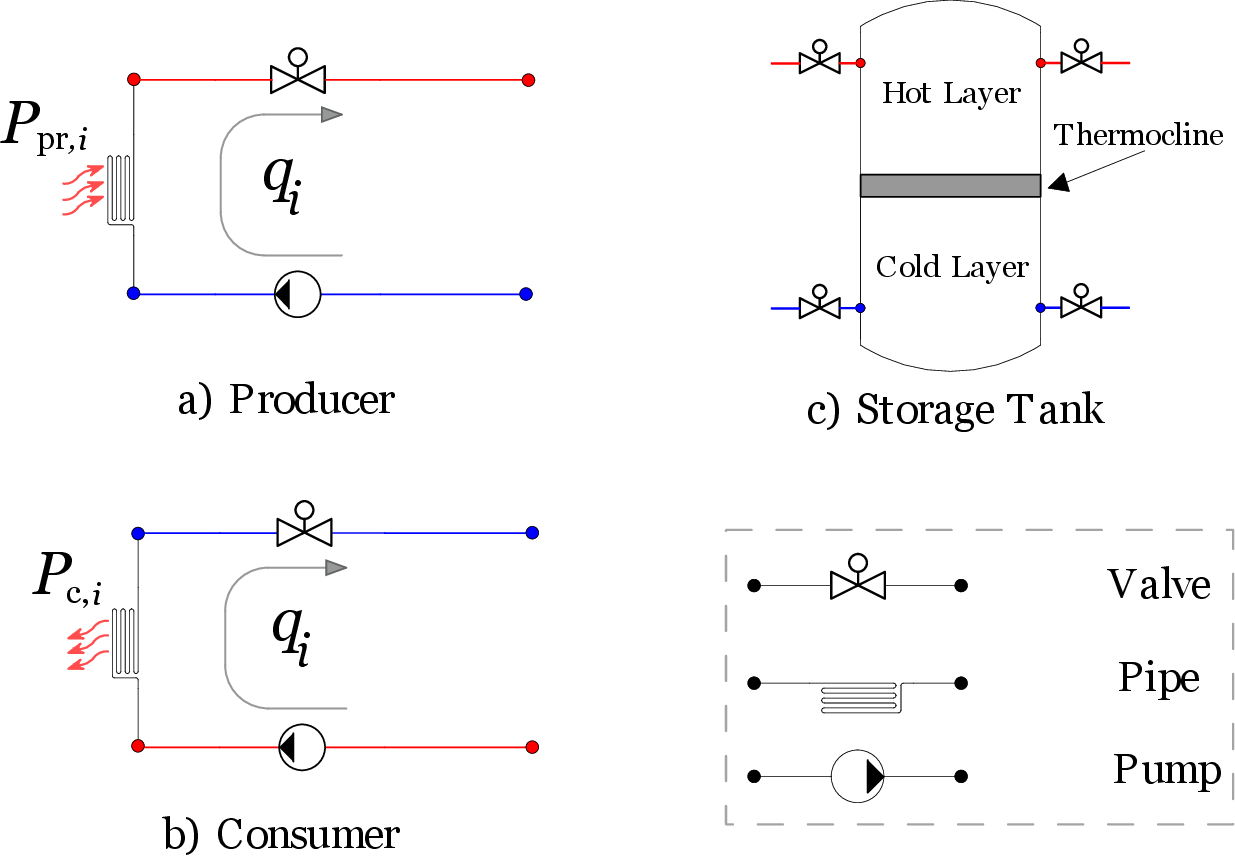}
\caption{Topologies of producers, consumers and storage tanks; see \cite{Scholten_tcst_2015,DePersis2011}.}
\label{fig:hydraulic_complex_elements}
\end{center}
\end{figure}

We consider a DH system with multiple heat producers, consumers and storage tanks. These devices are interconnected through a common meshed distribution network (see  Fig.~\ref{fig:dh_meshed}). Water is the medium used to transport thermal energy and we assume that temperatures and pressures in the system are such that water is always in liquid state.  The distribution network is considered to be conformed by two independent systems of pipelines. They respectively transport hot and cold water: hot water flows from heat producers to consumers via a supply layer and cold water flows in a converse manner through a return layer. It is assumed that the distribution network is symmetric, {\em i.e.,}  the supply and return layers are topologically equivalent, and that the physical interconnection between the two layers is done exclusively via  producers,  consumers or storage devices. The overall system is assumed to be leak free without any loss of the fluid mass \cite{talebi_review_16}. In this paper, producers, consumers and distribution network are conformed by basic hydraulic devices, namely, valves, pipes and pumps (see  Fig.~\ref{fig:hydraulic_complex_elements}). Next we provide a summary about the precise composition and operation mode of each producer, consumer and storage tank (see \cite{Scholten_tcst_2015} for more details). The hydraulic model of valves, pipes and pumps is detailed in Section~\ref{sec:hydraulic_model}.

\medskip

\revone{
\noindent {\em Producer (consumer):} Each producer (consumer) is represented as a connection in series of a pump, a pipe and a valve. The pipe models a tube through which the cold (hot) stream of a heat exchanger is flowing.  We assume that, via the exchanger, the producer (consumer) directly injects (draws) a thermal power $P_{{\mathrm{pr}},i}$ ($P_{\mathrm{c},i}$) into (out of) the DH system.
}

\medskip

\noindent {\em Storage tank:} A tank stores a mixture of hot and cold water which is perfectly separated by a  thermocline. The hot layer  is on top and the cold one at the bottom. It is assumed that there is no heat \revone{or mass} exchange between the mixtures.  The device has four valves, two at the top and two at the bottom, which are used as inlets and outlets of hot and cold water, respectively. 

%

\medskip

The overall DH system is viewed as a connected  graph $\mathcal{G}=(\mathcal{N},\mathcal{E})$ with  no self-loops (see, {\em e.g.},  \cite{DePersis2011,wang_meshed_17,
hauschild_ph_20,krug_DH_2021}). The set of edges $\mathcal{E}$ contains all  two-terminal devices (valves, pumps or pipes), and the set of nodes $\mathcal{N}$ contains all junctions as well as the hot and cold layers of each storage tank. The cardinalities of $\mathcal{N}$ and $\mathcal{E}$ are denoted by $n_{\mathrm N}$ and $n_{\mathrm E}$, respectively. Taking as reference the sketch in Fig.~\ref{fig:dh_meshed}, the gray, blue and red lines therein represent edges, whereas colored circles, nodes.

\revone{The variables $q_{\mathrm{E},i}$, $V_{\mathrm{E},i}$, $T_{\mathrm{E},i}$ and $p_{\mathrm{E},i}$ denote  the  flow rate, volume,  \revtwo{temperature and pressure} of the stream through $i\in \mathcal{E}$. Analogous descriptions follow for the variables $V_{\mathrm{N},k}$, $T_{\mathrm{N},k}$ and $p_{\mathrm{N},k}$, $k\in \mathcal{N}$.\footnote{\revtwoarxiv{A  summary of the most relevant variables and parameters appears in Appendix~\ref{app:summary_table}.}}

Also, we fix an arbitrary orientation to every edge of $\mathcal{G}$. Then, for any $i\in\mathcal{E}$ with end nodes $j,k\in \mathcal{N}$, $j\neq k$, we  say that $j$  is the head and $k$ is the tail of $i$, or viceversa, that $j$  is the tail and $k$ is the head of $i$. Then, for each node we  define the following sets  \cite{hauschild_ph_20} (see also \cite{krug_DH_2021}, \cite{valdimarsson_14}):
\begin{subequations}\label{eq:source_target_arbitrary_orien}
\begin{align}
\mathfrak{S}_k &  =  \{i\in \mathcal{E}: k~\text{is the tail of}~i\in \mathcal{E}\},~k\in \mathcal{N},\\
\mathfrak{T}_k  & =  \{i\in \mathcal{E}: k~\text{is the head of}~i\in \mathcal{E}\},~k\in \mathcal{N},
\end{align}
\end{subequations}
We define a constant incidence matrix $\mathcal{B}_0$ associated with the arbitrary orientation we have fixed for the DH system's edges, as follows:}
\begin{equation}\label{eq:incidence_matrix_B0}
{(\mathcal{B}_0)_{i,j}=}\resizebox{0.3\textwidth}{!}{$\revone{\begin{cases}
1\revtwo{,} & \text{if node $i$  is the head of edge  $j$},\\
-1\revtwo{,} & \text{if node $i$  is the tail of edge  $j$},\\
0\revtwo{,} & \text{otherwise}.
\end{cases}}$}
\end{equation}

\begin{remark}\label{rem:prelim_orientation}
\revone{For simplicity of exposition, we introduce  the preliminary assumption that the orientation of any edge $i\in \mathcal{E}$ matches the direction of the stream through it.  That is, if $j,k\in \mathcal{N}$, $j\neq k$, are the tail and head of any $i\in \mathcal{E}$, respectively,  then  the stream through $i$ is assumed to flow from $j$ to $k$ and, moreover, $q_{\mathrm{E},i}\geq 0$. However, some adjustments are necessary when modeling the DH system's temperature dynamics.}
\end{remark}

\begin{figure}
\begin{center}
\includegraphics[width=0.8\linewidth]{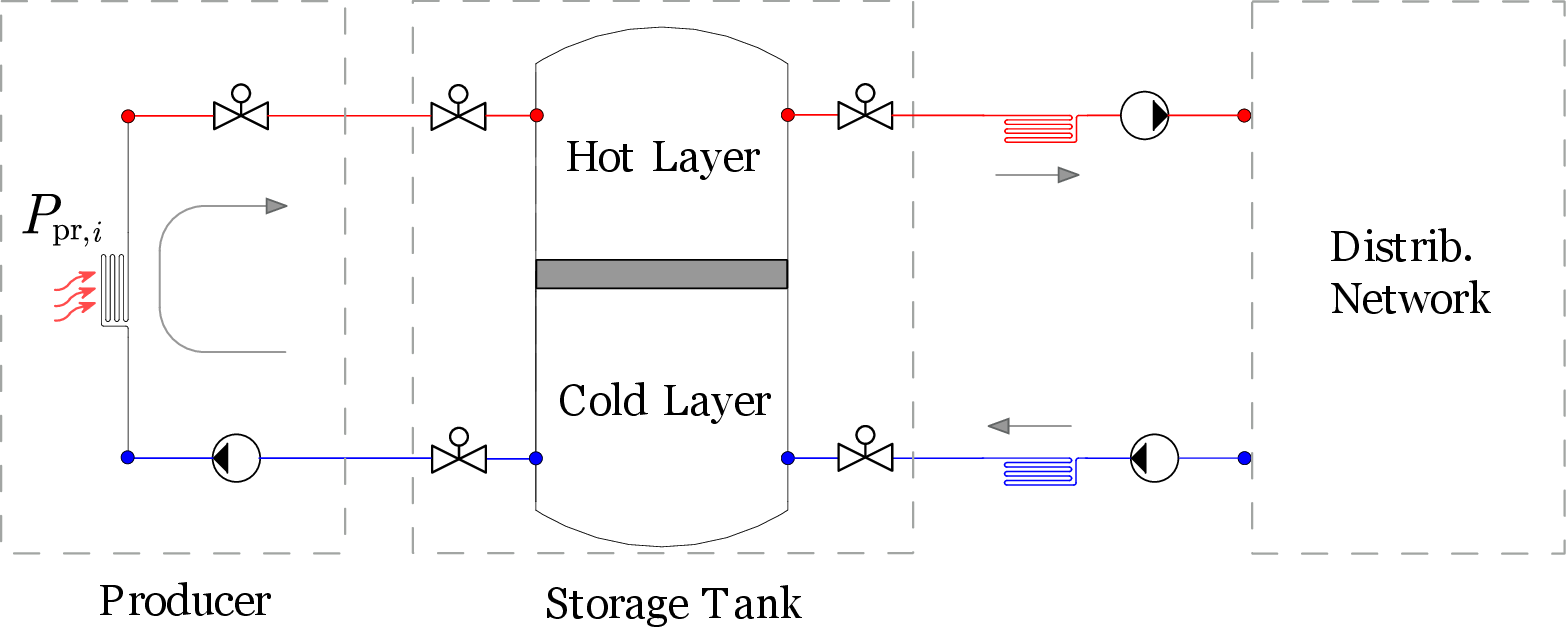}
\caption{Every producer in the system is interfaced to the distribution network through a storage tank. \revone{The gray arrows indicate the direction of the stream.}}
\label{fig:prod_stor}
\end{center}
\end{figure}

The following are standing assumptions in this work:

\begin{assumption}\label{assu:general_assumptions}
\revone{(i) The density $\rho>0$ and specific heat $c_{\mathrm{s.h.}}>0$ of water are spatially uniform and constant in time. (ii)~All pipes are cylindrical. (iii) The flow through any edge $i\in \mathcal{E}$ is (spatially) one-dimensional.  (iv)~Gravitational forces are neglected. (v) \revtwo{The pressure of each $k\in \mathcal{N}$ is spatially uniform and for each tank the pressure of its layers is equal.}} (vi) Each producer is interfaced to the distribution network through a storage tank  as depicted in Fig.~\ref{fig:prod_stor}. (vii) Each device (pipe, valve, pump, storage tank, junction) is completely filled with water all the time.  (viii)~There are no standalone storage tanks in the system.
\end{assumption}

\begin{remark}\label{rem:on the general assumptions}
\revone{
Assumptions~\ref{assu:general_assumptions}.(i)--(iii) are fairly common  in DH system modeling (see, {\em e.g.}, \cite{Scholten_tcst_2015,krug_DH_2021}).
%
%
Assumption~\ref{assu:general_assumptions}.(iv) is taken for simplification purposes.
%
We consider Assumption~\ref{assu:general_assumptions}.(v)  to discard inertial and viscous forces in the equations of  balance of momentum at each node and thus simplify the modeling procedure.
%
{\color{black} Assumption~\ref{assu:general_assumptions}.(vi) is taken without loss of generality. In fact, since producers  and consumers are comprised by the same type of devices and  have the same topology, then the modeling procedure and analysis described in this paper would still hold if we consider producers directly connected to the distribution network.} 
Assumption~\ref{assu:general_assumptions}.(vii) is also common and, in particular, it implies that for each  tank the sum of the volume  of the hot and cold layer is constant  (see,  {\em e.g.}, \cite{kamal_storage_97,Scholten_tcst_2015}). 
%
Assumption~\ref{assu:general_assumptions}.(viii) is mainly technical and  current work is underway to relax it.
}
\end{remark}

\begin{remark}
\revone{Storage tanks allow the continuous injection of heat to the system, at least for a period of time, even if the capacity of the associated heat producer is reduced. 
The design of control strategies for the adequate management of energetic resources in storage units is beyond the scope of this paper and is left as future research. Then, for the results we present in Section~\ref{sec:passivity_prop},   we assume that sensible equilibria of the DH system exist, {\em i.e.}, producers are considered to be able to supply all the  time the heat demanded by consumers.}
\end{remark}

\section{Hydraulic Model}\label{sec:hydraulic_model}

In this section we \revone{present} a model  to describe the dynamic  behavior of the  hydraulic variables of the DH system.

\revtwo{First, we present a  proposition  where we establish a hydraulic model written as a differential algebraic equation (DAE).  To streamline its presentation, we note that since the pressure is  uniform in each tank (see Assumption~\ref{assu:general_assumptions}.(v)), then we  find it convenient to reduce the  DH system's graph $\mathcal{G}$ by viewing  each tank as an individual node.  More precisely, let $\mathcal{N}_\mathrm{sh}$ and $\mathcal{N}_\mathrm{sc}$ denote the storage tanks' hot and cold layers', respectively, and $\mathcal{N}_\mathrm{sj}$ all the simple junctions. Then,  the graph $G=(N,\mathcal{E})$, with $N=\mathcal{N}_\mathrm{ST}\cup \mathcal{N}_\mathrm{sj}$,   $\mathcal{N}_\mathrm{ST}:=\mathcal{N}_\mathrm{sh}=\mathcal{N}_\mathrm{sc}$, represents a reduced version of $\mathcal{G}$ and $\vert N \vert=\vert \mathcal{N}\vert-n_\mathrm{ST}=:n_\mathrm{n}$. $G$ has the same edges as $\mathcal{G}$ and we assume they have the same orientation. Also, we associate to $G$  an incidence matrix $B_0$ which is analogous to \eqref{eq:incidence_matrix_B0}. With some abuse of notation, we also define sets analogous to \eqref{eq:source_target_arbitrary_orien}, and  use the same symbols. Then, by construction $G$ is connected.}

\revtwo{
\begin{proposition}\label{prop:hydraulic_DAE}
Let Assumption~\ref{assu:general_assumptions} hold. Let $p_\mathrm{n}$ and $V_\mathrm{n}$  denote the vector of pressures and volumes  of the nodes of $G$, respectively. Then, the overall flow and volume dynamics of the DH system is determined  by the following DAE:
\begin{subequations}\label{eq:hydraulic_model_dae_vector}
\begin{align}
-{B}_0 ^\top  p_\mathrm{n} & = \mathrm{diag}( J_{\mathrm{E}} ) \dot{q}_{\mathrm{E}}+f_{\mathrm{E}}(q_{\mathrm{E}})-w_{\mathrm{E}},\label{momentum_eq_vec}\\
0 & = {B}_0 q_\mathrm{E},\label{mass_bal_eq_vec}
\end{align}
\end{subequations}
where  $J_{\mathrm{E},i}= \rho \ell_{\mathrm{E},i}/{A_{\mathrm{E},i}}$ if $i\in \mathcal{E}$ is a pipe and zero otherwise, $f_{\mathrm{E},i}(q_{\mathrm{E},i})=K_{\mathrm{E},i} \vert q_{\mathrm{E},i}\vert q_{\mathrm{E},i}$, with $K_{\mathrm{E},i}>0$ if $i\in \mathcal{E}$ is either a pipe or a valve (and zero otherwise), and if $i\in \mathcal{E}$ is a pump, then $w_{\mathrm{E},i}$ is an external input representing the differential pressure it produces between its terminals, otherwise $w_{\mathrm{E},i}=0$. 
\end{proposition}
}

\revtwo{
\begin{pf}
We begin by establishing \eqref{momentum_eq_vec}. Let $i\in\mathcal{E}$ be an arbitrary pipe. Since $\rho$ is constant (see Assumption~\ref{assu:general_assumptions}.(i)), then  $q_{\mathrm{E},i}$ is a function of time only \cite{krug_DH_2021,hauschild_ph_20}. It follows that the momentum balance at $i$ can be written as \revtwoarxiv{(see Appendix~\ref{app:further_details_prop_flow_DAE} for details)}
\begin{align}\label{eq_rev30:press_drop}
p_{\mathrm{E},i}^\mathrm{in}-p_{\mathrm{E},i}^\mathrm{out} = J_{\mathrm{E},i}\dot{q}_{\mathrm{E},i}+K_{\mathrm{E},i} \vert q_{\mathrm{E},i} \vert q_{\mathrm{E},i},
\end{align}
where the second term in the right-hand side is the Darcy-Weisbach formula for frictional forces, and  $K_{\mathrm{E},i}=\theta_{\mathrm{E},i} \rho \ell_{\mathrm{E},i}/{2 D_{\mathrm{E},i}A_{\mathrm{E},i}^2}$ (see  \cite{hauschild_ph_20,grosswindhager_delay_11}).\footnote{We have assumed that the Darcy's friction factor $\theta_{\mathrm{E},i}$ is uniform along the pipe's axis and constant in time (see, {\em e.g.}, \cite{krug_DH_2021,rein_rom_2021}).}

Analogously, if $i\in \mathcal{E}$ is a valve, we  use  \eqref{eq_rev30:press_drop} to model the pressure drop through it.   Following \cite{DePersis2011}, we  neglect the inertial forces by taking $J_{\mathrm{E},i}=0$. Also  $K_{\mathrm{E},i}>0$ is in this case taken as  an unknown constant whose value depends on  generic   parameters, {\em e.g.},  the flow capacity and  opening degree of the valve \cite{wang_meshed_17}. 

Similarly, for any pump $i\in \mathcal{E}$ we model the pressure difference between inlet and outlet by $p_{\mathrm{E},i}^\mathrm{in}-p_{\mathrm{E},i}^\mathrm{out}=-w_{\mathrm{E},i}$, which is analogous to \eqref{eq_rev30:press_drop} (with $J_{\mathrm{E},i}=K_{\mathrm{E},i}=0$) and where  $w_{\mathrm{E},i}$ is the pressure difference produced by the pump.

From these developments, we can write the momentum balance equation at all edges in vector form as $p_{\mathrm{E}}^\mathrm{in}-p_{\mathrm{E}}^\mathrm{out} = \text{diag}(J_{\mathrm{E}})\dot{q}_{\mathrm{E}}+f_{\mathrm{E}}(q_{\mathrm{E}})-w_{\mathrm{E}}$; see \cite{DePersis2011}. To match the left-hand side of this equation with that of \eqref{momentum_eq_vec}, we bring the constraints $p_{\mathrm{E},i}^\mathrm{in}  = p_{\mathrm{n},k}$ and $p_{\mathrm{E},j}^\mathrm{out}  = p_{\mathrm{n},k}$, for all  $i\in \mathfrak{S}_k$, $j\in \mathfrak{T}_k$ and $k\in {N}$,  which guarantee that the pressure profile throughout \revtwo{the} DH system is continuous in space \cite{hauschild_ph_20} (see also \cite{krug_DH_2021,valdimarsson_14}). Then, if we use the incidence matrix, we get that $p_{\mathrm{E}}^\mathrm{in}-p_{\mathrm{E}}^\mathrm{out}=-B_0^\top p_\mathrm{N}$ (see \cite{valdimarsson_14}), which leads to~\eqref{momentum_eq_vec}.

To establish \eqref{mass_bal_eq_vec}, we note that in view of  Remark~\ref{rem:prelim_orientation}, the mass balance equation at any node $k\in{N}$ can be written as (see  \cite{hauschild_ph_20,krug_DH_2021,valdimarsson_14})
\begin{align}\label{eq:mass_balance_incidence}
\rho \dot{V}_{\mathrm{n},k}=\rho \sum_{i\in \mathfrak{T}_k}q_{\mathrm{E},i}-\rho \sum_{i\in \mathfrak{S}_k}q_{\mathrm{E},i},~k\in N.
\end{align}
According to Assumption~\ref{assu:general_assumptions}.(vii), the left-hand side of \eqref{eq:mass_balance_incidence} is zero  for all $k\in N$. Moreover, if we express it  in vector form, the right-hand side of the set  of all equations \eqref{eq:mass_balance_incidence} is equivalent to ${B}_0q_\mathrm{E}$. Therefore \eqref{mass_bal_eq_vec} is established.~$\hfill\blacksquare$

\end{pf}
}

\revone{Following \cite{DePersis2011},   we proceed to identify a set of independent flows from which the entire hydraulic state of the DH system can be determined. These flows are associated with a selected collection of {\em pipes}  that generate fundamental loops of $G$.

To be precise, we recall that $G$ is connected, then it has a (possibly non-unique) spanning tree (see~\cite{bollobas_book,DePersis2011}), which we denote by $S=(N,\mathcal{E}')$.}  Any $i\in \mathcal{E}\setminus \mathcal{E}'$ is referred to as  a {\em chord} of $S$ and the set of all  chords is denoted by $\mathcal{C}=\mathcal{E}\setminus \mathcal{E}'$. We denote by $n_{\mathrm{f}}=n_{\mathrm E} -(n_{\mathrm n}-1)$  the cardinality of $\mathcal{C}$.  Moreover, a fundamental  loop, denoted here by $\mathcal{L}_i$,  is the sequence of edges associated with the  loop that is formed  when  a chord is added to the spanning tree $S$.  It is assumed that each $\mathcal{L}_i$ has an orientation matching that of the chord that generates it. \revtwo{Then,  a {fundamental loop matrix}, denoted by $F\in\mathbb{R}^{n_{\mathrm{f}}\times n_{\mathrm E} }$, is  defined by components as $F_{i,h}=1$, if $h\in \mathcal{L}_i$ and orientations agree, $F_{i,h}=-1$, if $h\in \mathcal{L}_i$ and orientations disagree and $F_{i,h}=0$ if $h\notin \mathcal{L}_i$  (see \cite{DePersis2011}). We note that since $G$ is connected, then $F$ is a full-rank matrix (see \cite{DePersis2011}).}

\revtwo{In the considered DH system's setup, it can be verified that a suitable, yet not unique, selection of the set of chords is given by $\mathcal{C}=\mathcal{P}_\mathrm{c} \cup \mathcal{P}_\mathrm{pr}\cup \mathcal{P}_\mathrm{d}\cup \mathcal{P}_\mathrm{ST}$, where $\mathcal{P}_\mathrm{c}$ and $\mathcal{P}_\mathrm{pr}$ comprise all pipes associated with consumers and producers, respectively. Also, $\mathcal{P}_\mathrm{d}$ are all pipes that generate (fundamental) loops in the distribution network. Each element in $\mathcal{P}_\mathrm{ST}$ corresponds to the pipe at the hot water outlet of each storage tank (see Fig.~\ref{fig:prod_stor}), with the exception of one, which is chosen to belong to the spanning tree $S$ (see \cite{wang_meshed_17}). The latter consideration ensures that the spanning tree $S$ is  {\em connected} (see  \cite{DePersis2011}). In particular, we assume that the elements of  $\mathcal{P}_\mathrm{pr}$ and $\mathcal{P}_\mathrm{ST}$ are oriented towards the supply layer of the DH system. Conversely, the elements of $\mathcal{P}_\mathrm{c}$ are considered to point towards the return layer.}

We are in position to state this section's main result:
\begin{theorem}\label{theorem:overall_flows}
\revtwo{Let Assumption~\ref{assu:general_assumptions} hold. Also, let us assume there is an independently controlled pump adjacent to each $i\in \mathcal{C}$. Let $\mathcal{W}_\mathrm{f}\subset \mathcal{E}$ and $w_{\mathrm{f},i}$ denote the set of these pumps and the pressure difference produced  by each of them, respectively. If the  orientation of each pump in $\mathcal{W}_\mathrm{f}$ matches that of its adjacent chord, then the following holds true:}

\smallskip

\begin{enumerate}[label=(\Roman*), wide, labelwidth=!, labelindent=0pt]

\item The overall DH system's flows are determined by the flows through the chords. More precisely,  $q_\mathrm{E}=F^\top q_\mathrm{f}$, where $q_\mathrm{f}\in \mathbb{R}^{n_\mathrm{f}}$ is the vector of chords' flows.

\smallskip

\item All solutions of \eqref{eq:hydraulic_model_dae_vector} are determined by the ODE
\begin{equation}\label{eq:flow_dynamics_qf}
\mathcal{J}_\mathrm{f}\dot{q}_\mathrm{f}=\revone{-}f_\mathrm{f}(q_\mathrm{f}) \revone{+w_\mathrm{f}+B_\mathrm{b}w_\mathrm{b}},
\end{equation}
where $\mathcal{J}_f=F \text{diag}(J_\mathrm{E}) F^\top>0$, $f_\mathrm{f}(q_\mathrm{f})=Ff_\mathrm{E}(F^\top q_\mathrm{f})$. Also,  \revone{$B_\mathrm{b}w_\mathrm{b}$, with $(B_\mathrm{b})_{\alpha,\beta}\in \{-1,0,1 \}$, codifies the effect on $q_\mathrm{f}$ of any other pump} \revtwo{$i\in \mathcal{E}\setminus \mathcal{W}_\mathrm{f}$.}

\smallskip

\item \revtwo{There exists $W\in \mathbb{R}^{n_{\mathrm{ST}}\times n_\mathrm{f}}$, with entries in   $\{0,1\}$, such that the dynamics of the storage tanks' volumes are given by $\dot{V}_\mathrm{sh}=Wq_\mathrm{f}$ and $\dot{V}_\mathrm{sc}=-Wq_\mathrm{f}$, for the hot and cold layers, respectively.}

\end{enumerate}

\end{theorem}

\begin{pf}
\revtwo{The proof procedure of claims (I) and (II) is akin to the developments in \cite{DePersis2011}, but we present some details out of completeness.} Claim {(I)} is directly established by invoking the analogous versions of Kirchhoff's voltage and current laws for hydraulic networks. Following \cite{DePersis2011} (see also \cite{desoer_book}), they can be respectively stated as  
\begin{equation}\label{eq:kirchoff_laws}
\revtwo{F B_0^\top  p_\mathrm{n}}   = 0,~~~q_\mathrm{E}  = F^\top q_\mathrm{f}.
\end{equation}

These expressions represent also the starting point to establish claim (II). Indeed,  by combining \eqref{eq:kirchoff_laws}  with   \eqref{momentum_eq_vec} we obtain
\begin{subequations}\label{eq:dyn_proof}
\begin{align}
 F \text{diag}(J_{\mathrm E})F^\top \dot{q}_{\mathrm{f}}  & = -F f_\mathrm{E}(F^\top q_\mathrm{f})+F w_{\mathrm E}.
\end{align}
\end{subequations}
Moreover,   \eqref{mass_bal_eq_vec}, which represents the mass balance equations at the nodes of $G$, holds too. Indeed, both equations in \eqref{eq:kirchoff_laws} imply that  $B_0q_\mathrm{E}=B_0 F^\top q_\mathrm{f}=0$.

\revtwo{We note that \eqref{eq:dyn_proof}  matches \eqref{eq:flow_dynamics_qf}, except for  the terms related to the pumps' pressures. To see that $Fw_\mathrm{E}=w_\mathrm{f}+B_\mathrm{b}w_\mathrm{b}$, it is more convenient to assume, without loss of generality, that $\mathcal{E}$ is ordered as $\mathcal{C}\cup \mathcal{W}_\mathrm{f}\cup \mathcal{E}''$, where $\mathcal{E}''$ comprises the  remaining pipes, pumps and valves. This has two implications. On the one hand, by recalling that the edges in $\mathcal{C}$ and $\mathcal{W}_\mathrm{f}$ have the same orientation, then $F$ can be split as
\begin{equation}\label{eq:split_fund_loop_mat}
F=\left[I_{n_{\mathrm{f}}}~I_{n_{\mathrm{f}}}~B_\mathrm{b}\right],
\end{equation}
where $B_\mathrm{b}$ is as in the theorem's statement. On the other hand, $w_\mathrm{E}=\begin{bmatrix} \boldsymbol{0}_{n_\mathrm{f}}^\top,~w_\mathrm{f}^\top,~w_\mathrm{b}^\top \end{bmatrix}^\top$, which proves our assertion.}

\revtwo{
Regarding claim (III), let us assume that the nodes $\mathcal{N}$ of the primal DH system's graph $\mathcal{G}$ are ordered as $\mathcal{N}_\mathrm{sh}\cup \mathcal{N}_\mathrm{sc}\cup \mathcal{N}_\mathrm{sj}$. Then, the set of mass balance equations at each node in $\mathcal{N}$ is equivalent to $\dot{V}_\mathrm{N}=\mathcal{B}_0q_\mathrm{E}$, which combined with \eqref{eq:kirchoff_laws} produces $\dot{V}_\mathrm{N}=\mathcal{B}_0 F^\top q_\mathrm{f}$.  Pre-multiplying both sides of this equation by $W'=\begin{bmatrix} I_{n_\mathrm{ST}}, & 0_{n_\mathrm{ST}\times (n_\mathrm{N}-n_\mathrm{ST})}\end{bmatrix}$ leads to $\dot{V}_\mathrm{sh}=Wq_\mathrm{f}$, where $W=W'\mathcal{B}_0 F^\top$. To see that $\dot{V}_\mathrm{sc}=-Wq_\mathrm{f}$, we recall that \eqref{mass_bal_eq_vec},  which holds due to \eqref{eq:kirchoff_laws},   implies that $\dot{V}_\mathrm{sh}+\dot{V}_\mathrm{sc}=0_{n_\mathrm{ST}}$, {\em i.e.}, the sum of the volume of the hot and cold layer of each tank is constant (see Assumption~\ref{assu:general_assumptions}.(vii) and Remark~\ref{rem:on the general assumptions}).\footnote{\revtwoarxiv{More details about this proof appear in Appendix~\ref{app:further_details_proof_theorem_flows}.}}}~$\hfill\blacksquare$

\end{pf}

\begin{remark}\label{rem:ide_chords_pumps}
\begin{enumerate}[label=(\roman*), wide, labelwidth=!, labelindent=0pt]
\item Concerning the set of pumps $\mathcal{W}_\mathrm{f}$, we note that it is common that independently controlled pumps are placed at heat producers  \cite{wang_meshed_17},  consumers  \cite{DePersis2011,gong_multisource_2019}, and storage units \cite{guelpa_storage_review_2019}. Although some consumers or storage units might alternatively have control valves  \cite{wang_meshed_17,gong_multisource_2019},  in this work we restrict \revtwo{ourselves} to the multi-pump case only.  Moreover, we imply that there is a pump adjacent to each element in $\mathcal{P}_\mathrm{d}$. In   \cite{wang_meshed_17}, there are no control valves (nor pumps)  placed at these locations,  where the elements in $\mathcal{P}_\mathrm{d}$ are referred to as {\em residual branches}. Thus,  we are currently exploring the full extent  of the implications of relaxing this assumption in our work.

\smallskip

\item Costs associated with the system operation, which are related to pump power, may change in accordance to which setup and scheme is used to control system pumps and valves  \cite{wang_meshed_17,gong_multisource_2019}.  Such analysis is beyond the scope of our manuscript and is left as future research.    

\smallskip

\item Following \cite{DePersis2011}, we refer to any other pump in the system as a  {\em booster pump}. For simplicity we  assume that each of them produces  a constant pressure difference across its terminals; see \cite{DePersis2011} for more details.

\end{enumerate}

\end{remark}

\begin{remark}\label{rem:diff_claudios_etc}
\revone{
\begin{enumerate}[label=(\roman*), wide, labelwidth=!, labelindent=0pt]

\item \revtwo{Claim (III) of Theorem~\ref{theorem:overall_flows} implies that the total mass in the  DH system remains constant all the time. Thus, tanks will operate at full capacity provided this condition is met at the initial time instant. Critically, this does not guarantee that the volumes of the hot and cold layer of any storage are always simultaneously positive nor constant. Henceforth, we assume that the volume of each layer of each storage tank is positive all the time. This sensible assumption  is needed to guarantee that $\mathrm{diag}( V_\mathrm{th})$ in Theorem~\ref{theorem:shifted_pass_TDYN} is positive definite.}

\smallskip

\item Considering  Assumption~\ref{assu:general_assumptions},  it can be  shown that two decoupled dynamical systems can be obtained from \eqref{eq:flow_dynamics_qf}, one for the flow vector of  $\mathcal{P}_\mathrm{pr}$ and the  other for the flow vector of $\mathcal{P}_\mathrm{c}\cup \mathcal{P}_\mathrm{d}\cup \mathcal{P}_\mathrm{ST}$. \revtwoarxiv{More details can be found in Appendix~\ref{app:a_decoupling}.}

\smallskip

\item \revtwo{In the single  producer case treated in \cite{DePersis2011}, the orientation of the edges can be arranged such that a matrix analogous to  $B_\mathrm{b}$ in \eqref{eq:split_fund_loop_mat} has components only in $\{0,1\}$. This cannot be guaranteed in our system setup (multi-producer, meshed distribution network).
\revtwo{One salient implication of this is that  (component-wise) positive solutions $q_{\mathrm{f}}$ of~\eqref{eq:flow_dynamics_qf} do  not imply that $q_{\mathrm{E}}$ in \eqref{eq:hydraulic_model_dae_vector} remains in the positive orthant $\mathbb{R}^{n_{\mathrm{E}}}_{\geq 0}$.  In the next section, this is taken into account when defining a matrix which is central in the DH system's thermal dynamic modeling (see equation~\eqref{eq:flow-dep-incidence}).}}

\smallskip

\item Differently from Theorem~\ref{theorem:overall_flows}, the DH system's flow dynamics in \cite{hauschild_ph_20,krug_DH_2021,rein_rom_2021} are not formulated as an ODE.

\end{enumerate}}
\end{remark}

\section{Thermal Model}\label{sec:thermal_dyn}

In this section we present a model that describes the dynamic  behavior of the overall DH system's temperatures. Standing assumptions behind this model, which complement those in Assumption~\ref{assu:general_assumptions},  are the following ({\em c.f.}, \cite{VanderHeijde2017,kamal_storage_97,Scholten_tcst_2015}):
\begin{assumption}\label{assu:thermal_model}
(i)~Kinetic, potential and flow energies are negligible compared to the internal energy of water.  (ii)~Heat conduction  is neglected. (iii)~Dissipation due to friction is  negligible. (iv)~The internal energy of water linearly depends on the temperature. (v)~The rate at which work is done by external forces that act on the (moving) boundary of any tank's layer  is negligible. \revtwo{(vi)~The temperature $T_{\mathrm{E},i}$ ($T_{\mathrm{N},k}$) of each $i\in \mathcal{E}$  ($ k \in \mathcal{N}$) represents a  spatially averaged quantity  over its control volume.}
\end{assumption}

\revtwo{
We note that Assumption~\ref{assu:thermal_model}.(i)--(iv) holds for all edges and nodes, while  Assumption~\ref{assu:thermal_model}.(v) only for the layers of the storage tanks.  Additionally,  by neglecting  heat conduction,  the heat transfer through the boundaries between edges and nodes is discarded (as well as the heat transfer towards the environment); for storage tanks this also means that there is no heat transfer through the thermocline.  Also,  Assumption~\ref{assu:thermal_model}.(v) is introduced to take into consideration the fact that if $k$ represents a layer of a storage tank, then $\dot{V}_{\mathrm{N},k}$ is not necessarily zero  all the time, {\em i.e.}, its boundary (the thermocline) might move. Then,  when we write later the energy balance at $k$, which is a deforming control volume (see, {\em e.g.}, \cite{sonin_deformable_CV_2001}), we discard the  effects of the boundary forces;  volume-varying nodes are not considered in \cite{hauschild_ph_20,krug_DH_2021,rein_rom_2021}. Assumption~\ref{assu:thermal_model}.(vi) may bring inaccuracies in the description of the temperature profile of long pipes. Note however that in  such cases a pipe can be viewed as a series connection of shorter pipes of similar properties; see \cite{hauschild_ph_20} for a rigorous treatment on spatial discretization of PDE-based pipe models.
}

\subsection{Edges}

\revtwo{Let Assumptions~\ref{assu:general_assumptions} and \ref{assu:thermal_model} hold and let $i\in \mathcal{E}$ be an arbitrary pipe of the DH system.}  For simplicity of exposition, we assume that the orientation of  $i$ matches the direction of the stream through it, then,  $q_{\mathrm{E},i}\geq 0$ all the time.  
%
%
%
Then, the energy balance at $i$ can be written  as follows:
\begin{align}\label{eq:model_pipe_no_heat_trans}
\rho c_{\mathrm{s.h.}} V_{\mathrm{E},i}\revtwo{\dot{T}_{\mathrm{E},i}}=\rho  c_{\mathrm{s.h.}}q_{\mathrm{E},i}\left(T_{\mathrm{E},i}^\mathrm{in}-T_{\mathrm{E},i}^\mathrm{out}\right),
\end{align}
\revtwo{where $T_{\mathrm{E},i}^\mathrm{in}$ ($T_{\mathrm{E},i}^\mathrm{out}$) is the temperature  at the inlet (outlet) of the pipe.}

Based on \eqref{eq:model_pipe_no_heat_trans}, we model the heat exchangers (which are viewed as pipes) at  producers and consumers as follows  \cite{Scholten_tcst_2015}
\begin{align}\label{eq:app_pipe_prod_cons}
\rho c_{\mathrm{s.h.}} V_{\mathrm{E},i}  \dot{T}_{\mathrm{E},i}  =  \rho c_{\mathrm{s.h.}} q_{\mathrm{E},i} \left(T_{\mathrm{E},i}^\mathrm{in}-T_{\mathrm{E},i}^\mathrm{out} \right)+P_{{\mathrm{pr}},i}-P_{{\mathrm c},i},
\end{align}
where $P_{{\mathrm{pr}},i}$ is the thermal power the $i$th producer transfers into the pipe's stream and $P_{{\mathrm c},i}$  is the thermal power the consumer extracts from it.  If $i\in \mathcal{E}$ is associated with a producer, we take $P_{{\mathrm c},i}=0$, and  $P_{{\mathrm{pr}},i}=0$ if  $i$ is associated with a consumer.

\begin{remark}

\begin{enumerate}[label=(\roman*), wide, labelwidth=!, labelindent=0pt]

\item We also use the  model  \eqref{eq:app_pipe_prod_cons} to describe the thermal behavior of all valves and pumps in the DH  system. Then,  \eqref{eq:app_pipe_prod_cons}  describes the temperature dynamics of any edge $i\in \mathcal{E}$. \revtwo{Note that $P_{\mathrm{pr},i}$ and $P_{\mathrm{c},i}$ are associated with producers and consumers' pipes, then we take them as zero for any other type of edge.}

\smallskip

\item In the following section we describe the thermal dynamics of the DH system's nodes. Therein, we  write a relationship between  the temperature $T_{\mathrm{E},i}^\mathrm{in}$ at the inlet of any edge $i$ with the temperature of the node from which the stream through $i$ sources  from. 

\end{enumerate}

\end{remark}

\subsection{Nodes}

\revtwo{Again, let Assumptions~\ref{assu:general_assumptions} and \ref{assu:thermal_model} hold and let  $k\in \mathcal{N}$ be an arbitrary node of the DH system.} For simplicity we assume again and without loss of generality that the orientation of any edge $i\in \mathcal{E}$ matches the direction of the stream through it, {\em i.e.}, we assume that $q_{\mathrm{E},i}\geq 0$ (see  Remark~\ref{rem:thermal_flows_wlog}).
Considering Assumptions~\ref{assu:general_assumptions} and \ref{assu:thermal_model},  it is possible to write the energy balance at any $k\in \mathcal{N}$ as follows  ({\em c.f.}, \cite{hauschild_ph_20,krug_DH_2021}):
\begin{align}\label{eq:thermal_model_detail_nodes_0}
\frac{\mathrm{d}}{\mathrm{d} t}\left(\rho  c_{\mathrm{s.h.}} V_{\mathrm{N},k} T_{\mathrm{N},k}\right) & = \sum_{j\in \mathfrak{T}_k} \rho c_{\mathrm{s.h.}} q_{\mathrm{E},j}T_{\mathrm{E},j}^{\mathrm{out}} \nonumber \\
& ~~~~ - \sum_{j\in \mathfrak{S}_k} \rho c_{\mathrm{s.h.}} q_{\mathrm{E},j}T_{\mathrm{E},j}^\mathrm{in}.
\end{align}
The term in the left-hand side represents the thermal energy stored at node $k$ whereas  the terms in the right-hand side are the sum of the thermal energies of the streams that target and source from $k$, respectively.

\revtwo{Based on the (upwind)  semi-discretization scheme discussed in\cite{hauschild_ph_20} and on the nodal constraints described in  \cite{krug_DH_2021},  we impose on \eqref{eq:app_pipe_prod_cons} and \eqref{eq:thermal_model_detail_nodes_0}  the following conditions:}
\begin{subequations}\label{eq:tempe_node_edge}
\begin{align}
T_{\mathrm{E},j}^\mathrm{in} & =T_{\mathrm{N},k},~j\in \mathfrak{S}_k,~k\in \mathcal{N},\\
\revtwo{T_{\mathrm{E},j}^\mathrm{out}} & = \revtwo{T_{\mathrm{E},j},~j\in \mathcal{E}.}
\end{align}
\end{subequations}
\revtwo{In particular, these conditions allow us to write \eqref{eq:thermal_model_detail_nodes_0}  in a simpler, equivalent form. Indeed, since   the mass balance equation at node $k\in \mathcal{N}$ can be written as  $\rho \dot{V}_{\mathrm{N},k}=\sum_{j\in \mathfrak{T}_k} \rho q_{\mathrm{E},j}- \sum_{j\in \mathfrak{S}_k} \rho q_{\mathrm{E},j}$, then \eqref{eq:thermal_model_detail_nodes_0} is equivalent to}
\begin{align}\label{eq:thermal_model_detail_nodes}
\rho c_{\mathrm{s.h.}} V_{\mathrm{N},k} \dot{T}_{\mathrm{N},k} & = \sum_{j\in \mathfrak{T}_k} \rho c_{\mathrm{s.h.}} q_{\mathrm{E},j}T_{\mathrm{E},j} \nonumber \\
& ~~~~ - \left( \sum_{j \in \mathfrak{T}_k} \rho c_{\mathrm{s.h.}}q_{\mathrm{E},j}\right) T_{\mathrm{N},k},~k\in \mathcal{N}.
\end{align}
For simplicity we henceforth let $\rho=c_{\mathrm{s.h}}=1$.

\begin{remark}\label{rem:thermal_flows_wlog}

\begin{enumerate}[label=(\roman*), wide, labelwidth=!, labelindent=0pt]


\medskip

\item The models \eqref{eq:app_pipe_prod_cons}--\eqref{eq:thermal_model_detail_nodes} need to be  adjusted if the orientation of a given edge $i\in \mathcal{E}$ does not match the direction of the stream through it, {\em i.e.}, if $q_{\mathrm{E},i}(t)\leq 0$ for certain $t$. In this case it is necessary to substitute $q_{\mathrm{E},i}$ by its absolute value and re-define the sets $\mathfrak{S}_k$ and $\mathfrak{T}_k$ as follows  \cite{hauschild_ph_20,krug_DH_2021}, ({\em c.f.}, \cite{valdimarsson_14})
\begin{subequations}\label{eq:sets_frakS_frakT}
\begin{align}
\mathfrak{S}_k & =  \{i\in \mathcal{E}: \left(k~\text{is the tail of}~i~\text{and}~q_{\mathrm{E},i}\geq 0 \right) \nonumber  \\
&~~~~~~ ~\text{or}~  \left(k~\text{is the head of}~i~\text{and}~q_{\mathrm{E},i}< 0 \right)\},\\
\mathfrak{T}_k & =  \{i\in \mathcal{E}: \left(k~\text{is the tail of}~i~\text{and}~q_{\mathrm{E},i}< 0 \right) \nonumber \\
& ~~~~~~~~\text{or}~  \left(k~\text{is the head of}~i~\text{and}~q_{\mathrm{E},i}\geq  0 \right)\}.
\end{align}
\end{subequations}
In the sequel we say that node  $k$ is the source of  the stream through any $i\in\mathfrak{S}_k$ and that  $k$ is the target of the stream through any  $i\in\mathfrak{T}_k$. 

\end{enumerate}

\end{remark}


\subsection{Thermal model in vector form}

To write in vector form the system of equations \eqref{eq:app_pipe_prod_cons}, \eqref{eq:tempe_node_edge} and \eqref{eq:thermal_model_detail_nodes},   we find it convenient to introduce a flow-dependent, node-edge incidence matrix. We denote this matrix by $\mathcal{B}(q_{\mathrm E})\in \mathbb{R}^{n_{\mathrm N}\times n_{\mathrm E}}$ and define it as follows: $\mathcal{B}_{ij}(q_\mathrm{E})=1$, if the flow through $j\in \mathcal{E}$ targets $i\in \mathcal{N}$; $\mathcal{B}_{ij}(q_\mathrm{E})=-1$, if the flow through $j\in \mathcal{E}$ originates from $i\in \mathcal{N}$; and $\mathcal{B}_{ij}(q_\mathrm{E})=0$, otherwise. Following \cite{valdimarsson_14}, this matrix can be compactly  written as
\begin{align}\label{eq:flow-dep-incidence}
\mathcal{B}=\mathcal{B}_0 \mathrm{diag}( \mathbf{sign}(q_{\mathrm E}) ),
\end{align}
where $\mathcal{B}_0$ is given in \eqref{eq:incidence_matrix_B0}. \revone{To our purposes,  the positive and negative parts of $\mathcal{B}$ \revtwo{are more relevant}, since they  respectively identify (for each edge) its target and source nodes (see \cite{valdimarsson_14}). These matrices are respectively defined as follows (see \cite{valdimarsson_14}):
\begin{align}\label{eq:incidence_target}
\mathcal{T}  =\frac{1}{2}\left( \mathcal{B}+\vert \mathcal{B} \vert  \right),~~\revone{\mathcal{S}  = \frac{1}{2}\left(\vert \mathcal{B}\vert- \mathcal{B} \right)}.
\end{align}
We note that $\mathcal{B}$, $\mathcal{T}$ and $\mathcal{S}$ are  time-varying  in general.
}

Consider then the following theorem (its proof  can be established via direct computations and is therefore omitted for brevity).
\begin{theorem}\label{thm:thermal_matrix_primal}
The system \eqref{eq:app_pipe_prod_cons}, \eqref{eq:tempe_node_edge} and \eqref{eq:thermal_model_detail_nodes}  can be equivalently represented in vector form as follows:
\begin{equation}\label{eq:dyntemp_nonred}
\mathrm{diag}(V_\mathrm{E},V_\mathrm{N})\begin{bmatrix}
 \dot{T}_{\mathrm E}\\
\dot{T}_{\mathrm N}
\end{bmatrix}  = \mathcal{A}(q_\mathrm{E}) \begin{bmatrix}
T_{\mathrm E}\\
T_{\mathrm N}
\end{bmatrix}  + \begin{bmatrix} \mathbf{P}_{\mathrm{pr}} -\mathbf{P}_{\mathrm c}\\
 0_{n_\mathrm{N}} \end{bmatrix},
\end{equation}
where 
\begin{equation}\label{eq:calA}
\mathcal{A}(q_\mathrm{E})  =\begin{bmatrix}
- \mathrm{diag}( \vert q_{\mathrm E}  \vert ) &  \mathrm{diag}( \vert q_{\mathrm E} \vert) \mathcal{S}^\top\\
\mathcal{T} \mathrm{diag} (\vert q_{\mathrm E} \vert ) & - \mathrm{diag}( \mathcal{T} \vert q_{\mathrm E} \vert),
\end{bmatrix}
\end{equation}
and $\mathbf{P}_{\mathrm{pr}}=[P_{{\mathrm{pr}},1},\dots,P_{{\mathrm{pr}},n_{\mathrm E}}]^\top$ and $\mathbf{P}_{\mathrm c}=[P_{{\mathrm c},1},\dots,P_{{\mathrm c},n_{\mathrm E}}]^\top$. 
\end{theorem}
%

The model \eqref{eq:dyntemp_nonred} represents an ODE and is nonlinear due to the dependency of $\mathcal{A}$ on  $q_\mathrm{E}$, and  due to the time-varying behavior of the \revtwo{volumes} associated with the layers of the storage tanks.

\revtwo{
We note that due to the small volume of valves, pumps and junctions, it is reasonable to assume they  have a negligible thermal inertia compared to pipes and storage tanks. If we take  $V_{\mathrm{E},i}=0$ for any $i\in \mathcal{E}$ associated with a valve or pump, and  $V_{\mathrm{N},j}=0$ for any $j\in \mathcal{N}$ associated with a simple junction, then thermal inertia of these devices is neglected. This reduces the number of differential variables of \eqref{eq:dyntemp_nonred}, but turns it into a system of semi-explicit DAEs. \revtwoarxiv{In Appendix~\ref{app:diff_index_1_noncontractedgraph} we discuss an  assumption for  this system to have differentiation index 1, which facilitates the computation of its trajectories. However, its verification is  not direct  in general.}}
\revtwo{
An alternative approach  consists in defining a reduced DH system's graph $\tilde{\mathcal G}=(\tilde{\mathcal N},\tilde{\mathcal E})$ obtained by {\em contracting} all the edges associated with valves and pumps in $\mathcal{G}$, provided they are assumed to have zero volume.  To prevent that $\tilde{\mathcal G}$ has self-loops, we make the following, practically sensible assumption:

\begin{assumption}\label{assu:model_reduc}
Each  valve and pump of the overall DH system's graph  $\mathcal{G}$ is in series with a pipe. 
\end{assumption}
}

\revtwo{
Let $q_\mathrm{r}$ denote the flows through the pipes of $\tilde{\mathcal{G}}$. Thus, by virtue of Assumption~\ref{assu:model_reduc}, the components of $q_\mathrm{r}$ are those of $q_\mathrm{E}$ in \eqref{eq:kirchoff_laws} associated with the (same) pipes in $\mathcal{G}$.  In the following corollary, whose proof is in \revtwoarxiv{Appendix~\ref{app:proof_corollary_rom}},  any variable, set or matrix  defined for $\mathcal{G}$ and the system \eqref{eq:dyntemp_nonred}  ({\em e.g.}, the matrix $\mathcal{A}$, or the sets $\mathfrak{S}_k$ and $\mathfrak{T}_k$) is analogously defined for $\tilde{\mathcal G}$ and identified with the symbol $(\tilde{\cdot})$.
}

\begin{corollary}\label{corollary: thermal_state_dyn}
\revtwo{Let Assumptions~\ref{assu:general_assumptions}--\ref{assu:model_reduc} hold and assume that all valves, pumps and simple junctions in the DH system have zero volume.  Let $\tilde{\mathcal{N}}_0\subset \tilde{\mathcal{N}}$ denote the set of simple junctions in $\tilde{\mathcal{G}}$ and assume that $\sum_{k  \in \tilde{\mathfrak{T}}_j} \vert \tilde{q}_{{\mathrm E},k}\vert >0$ for each $j\in \tilde{\mathcal{N}}_0$.} Then,   the state of the temperatures in all the system's pipes and storage tanks' layers  is determined by a vector  $T_{\mathrm{th}}$ \revone{comprising the temperatures of each $k\in \tilde{\mathcal{E}}\cup \left(\tilde{\mathcal{N}}\setminus \tilde{\mathcal{N}}_0 \right)$,} which  satisfies a dynamic equation of the form
\begin{align}\label{eq:TDYN_redu}
\mathrm{diag}( V_\mathrm{th} ) \dot{T}_{\mathrm{th}} = A_{\mathrm{th}}(q_\mathrm{r}) T_{\mathrm{th}} + \revone{B_{\mathrm{pr}} P_{\mathrm{pr}} - B_\mathrm{c}P_{\mathrm{c}}},
\end{align} 
\revone{where $A_{\mathrm{th}}(q_\mathrm{r})$ is the Schur complement of an appropriate  sub-matrix of $\tilde{\mathcal{A}}({q}_\mathrm{r})$, and $(B_{\mathrm{pr}})_{\alpha,\beta}, \in \{0,1 \}$ and $(B_{\mathrm{c}})_{\alpha,\beta}, \in \{0,1 \}$ are properly sized, full (column) rank matrices.}\footnote{\revtwoarxiv{An extension to this corollary is discussed in Appendix~\ref{app:extension_corollary_1}.}}
\end{corollary}
\begin{pf}
\revtwoarxiv{See Appendix~\ref{app:proof_corollary_rom}.~$\hfill\blacksquare$}
\end{pf}

\begin{remark}\label{rem:remarks_temps_dyn}
\revone{
\begin{enumerate}[label=(\roman*), wide, labelwidth=!, labelindent=0pt]

\item Henceforth we view each $P_{\mathrm{pr},i}$ as a control input and each $P_{\mathrm{c},i}$ as an external disturbance.


\smallskip

\item By directly working with \eqref{eq:dyntemp_nonred},  we avoid imposing conditions on the flows as we do to define \eqref{eq:TDYN_redu}. Besides the main drawback of having a higher order, some values of $V_{\mathrm{E},i}$ and $V_{\mathrm{N},j}$ in \eqref{eq:dyntemp_nonred} could be uncertain, as it could be difficult to accurately measure them  for all valves, pumps  and junctions in the DH system. Nonetheless, the shifted passivity properties that we present in Section~\ref{sec:passivity_prop}  for \eqref{eq:TDYN_redu}  can be analogously established for \eqref{eq:dyntemp_nonred}. Notably, knowledge about the system elements' volumes would not be necessary for that purpose, as is the case for the passivity-based controllers discussed in Remark~\ref{rem:shifted_passivity}.

\end{enumerate}}
\end{remark}

\section{Passivity Properties}\label{sec:passivity_prop}

In this  section we establish that the DH system's  flow and thermal dynamics are shifted passive  subject to the assumption that there is a distinct time scale separation between these subsystems. The plausibility of this hypothesis stems from the fact that the flows through the system 
 reach a steady-state in a matter of seconds, whereas temperature changes at the heat production sites need hours to affect consumer stations \cite[Section~2.4]{grosswindhager_delay_11}, \cite[Section~2.3]{palsson_book_99} and \cite[Section~2]{benonysson_opt_95}.

To improve the readability, consider the following.

\begin{definition}[\cite{arjan_l2gain_book}]\label{def:shifted_pass}
\revone{Consider a dynamic system $\Sigma: \dot{x}=f(x,u)$, with input $u$ and output $y=h(x,u)$, of the same size. Let $(\bar{x},\bar{u})$ denote an equilibrium pair of $\Sigma$ and define $\bar{y}=h(\bar{x},\bar{u})$. Then, $\Sigma$ is shifted passive if there exists a non-negative scalar function $\mathcal{H}(x)$ such that, along any  solution $x$ of $\Sigma$, it satisfies \revone{$\dot{\mathcal{H}}(x)\leq (u-\bar{u})^\top (y-\bar{y})$.}}
\end{definition}

\revone{In  Remark~\ref{rem:shifted_passivity} we discuss some of the benefits for control design and stability analysis of identifying that flow and thermal dynamics are shifted passive.\footnote{\revone{The definition of {\em passivity} follows from Definition~\ref{def:shifted_pass} by taking $\bar{u}=\bar{y}=\mathbf{0}$.}}
}

\subsection{Passivity of the Flow Dynamics}\label{sec:passivity_flow_dyn}

We first focus on the flow dynamics~\eqref{eq:flow_dynamics_qf}, which we remark are independent with respect to the DH system's temperatures.
Before presenting the main result of this subsection, consider the following lemma, where an important property of the mapping $f_{\mathrm{f}}$ is introduced (see Theorem~\ref{theorem:overall_flows}). The proof is omitted since it can be established by following the proof's procedure of \cite[Lemma 3.1]{DePersis2014}. \revone{Nonetheless, it is important to note that each \revtwo{mapping} $f_{\mathrm{E},i}(q_{\mathrm{E},i})$ in \eqref{eq:hydraulic_model_dae_vector} is a continuously differentiable monotonic function; the latter means that the following results would hold for any other friction model $f_{\mathrm{E},i}$ as long as it satisfies such properties.}

\begin{lemma}\label{lemma:-f_monotone}
The mapping $f_{\mathrm f}$ that appears in the right-hand side of the flow dynamics~\eqref{eq:flow_dynamics_qf} is monotone with respect to $q_{\mathrm f}$.\footnote{A mapping  $\mathcal{F}:\mathbb{R}^n\rightarrow \mathbb{R}^n$ is said to be monotone if 
$
(u-v)^\top (\mathcal F(u)- \mathcal F(v))\geq 0,~\forall u,v\in\mathbb{R}^n
$, \cite[Section 4]{boyd_monotone_16}.
In the case that $\mathcal F$ is differentiable then a necessary and sufficient condition for monotonicity is given by 
$
\nabla \mathcal F(x)+\nabla \mathcal F(x)^\top \geq 0,~\forall x\in\mathbb{R}^n 
$, \cite[p.12]{boyd_monotone_16}.}
\end{lemma}
%
We are in a position to prove the following theorem:
\begin{theorem}\label{theorem:shifted_passivity}
The DH system's flow dynamics~\eqref{eq:flow_dynamics_qf} are shifted passive with storage function $\mathcal{H}_{\mathrm f}(q_{\mathrm f})=\frac{1}{2}(q_{\mathrm f}-\bar q_{\mathrm f})^\top \mathcal{J}_{\mathrm f} (q_{\mathrm f}-\bar q_{\mathrm f})$ and shifted passive output $y_{\mathrm f}-\bar{y}_{\mathrm f} = q_{\mathrm f}-\bar q_{\mathrm f}$, where $(\bar q_{\mathrm f},\revone{\bar{w}_{\mathrm f}})$ is any equilibrium pair of the system.
\end{theorem}
\begin{pf}
Since $(\bar q_{\mathrm f},\revone{\bar w_{\mathrm f}})$ denotes any equilibrium pair of the flow dynamics~\eqref{eq:flow_dynamics_qf}, then the identity  $\mathcal{J}_{\mathrm f}\dot{\bar{q}}_{\mathrm f}=\boldsymbol{0}=-f_{\mathrm f}(\bar q_{\mathrm f})+\revone{\bar w_{\mathrm f}+B_\mathrm{b}w_\mathrm{b}}$ holds. Therefore, \eqref{eq:flow_dynamics_qf} can be equivalently written as follows (recall that $w_\mathrm{b}$ is constant):
\begin{align}\label{eq:shifted_q_f-dyn}
\mathcal{J}_{\mathrm f}\dot{q}_{\mathrm f}= -(f_{\mathrm f}( q_{\mathrm f})-f(\bar q_{\mathrm f}))+\revone{w_{\mathrm f}-\bar w_{\mathrm f}}.
\end{align}
Consider the mapping  $\mathcal{H}_{\mathrm f}$ as a candidate  storage function to prove that \eqref{eq:shifted_q_f-dyn} is shifted passive; clearly $\mathcal{H}_{\mathrm f}$  is continuously differentiable and bounded from below by zero and, moreover, along  the solutions of system~\eqref{eq:shifted_q_f-dyn}, its time derivative satisfies $ \dot{\mathcal{H}}_{\mathrm f}(q_{\mathrm f})   = \left(q_{\mathrm f}-\bar q_{\mathrm f}\right)^\top \left( -(f_{\mathrm f}(q_{\mathrm f})-f_{\mathrm f}(\bar q_{\mathrm f})) + \revone{w_{\mathrm f}-\bar w_{\mathrm f} }\right).$
From {Lemma~\ref{lemma:-f_monotone}}, the mapping $f_{\mathrm f}$ is monotone with  respect to $q_{\mathrm f}$, yielding that  $-(q_{\mathrm f}-\bar q_{\mathrm f})^\top (f_{\mathrm f}(q_{\mathrm f})-f_{\mathrm f}(\bar q_{\mathrm f})) \leq  0$. Consequently, $\dot{\mathcal{H}}_{\mathrm f} (q_{\mathrm f}) \leq (y_{\mathrm f}-\bar y_{\mathrm f})^\top (\revone{w_{\mathrm f}-\bar w_{\mathrm f}})$. $\hfill\blacksquare$
\end{pf}

\subsection{Passivity of the Thermal Dynamics}\label{subsec:passiv_thermal_dyn}

Here we focus  on the DH system's thermal dynamics~\eqref{eq:TDYN_redu}, \revtwo{however analogous results hold also  for \eqref{eq:dyntemp_nonred}, as it will be clear from the proofs of the following results.}

\revtwo{Let $\bar{q}_\mathrm{f}$ denote an equilibrium of \eqref{eq:flow_dynamics_qf}, and let us assume that the flow dynamics~\eqref{eq:flow_dynamics_qf} have a much faster response than the thermal model~\eqref{eq:TDYN_redu}.  Then, in \eqref{eq:TDYN_redu}  we can  take $q_{\mathrm r}=\bar q_{\mathrm r}$ all the time, where $\bar{q}_\mathrm{r}$ is an equilibrium flow depending on $\bar{q}_\mathrm{f}$ (see the definition of $q_\mathrm{r}$ just above Corollary~\ref{corollary: thermal_state_dyn}).}


\begin{lemma}\label{lem:calA_neg_sem}
The coefficient matrix $A_\mathrm{th}(q_\mathrm{r})$ of the thermal dynamics \eqref{eq:TDYN_redu} is negative semi definite at any  equilibrium $\bar{q}_\mathrm{r}$.
\end{lemma}
\begin{pf}
\revtwo{Let us begin by showing that $\tilde{\mathcal{A}}(\bar{q}_\mathrm{r})$ is a Kirchhoff Convection Matrix (KCM) and thus negative semi definite (see  \cite[Appendix]{hangos_thermo_99}). We recall  that $\tilde{\mathcal A}$ is analogous to $\mathcal{A}$ in \eqref{eq:calA}, but defined for the reduced graph $\tilde{\mathcal{G}}$ instead of $\mathcal{G}$.}

Let us treat the case in which each component of $\bar{q}_\mathrm{r}$ is different from zero. Following \cite[Appendix]{hangos_thermo_99},  \revtwo{$\tilde{\mathcal{A}}(\bar{q}_\mathrm{r})$ is a KCM if and only if the following four conditions hold: (i)  $\tilde{\mathcal{A}}_{i,i}(\bar q_{\mathrm r})<0$; (ii) $\tilde{\mathcal{A}}_{i,j}(\bar q_{\mathrm r})\geq 0$; (iii) $\tilde{\mathcal{A}}(\bar q_{\mathrm r})\boldsymbol{1}=\boldsymbol{0}$; and $\boldsymbol{1}^\top\tilde{\mathcal{A}}(\bar q_{\mathrm r})=\boldsymbol{0}^\top$.  Condition (i) holds since each $\tilde{\mathcal{A}}_{i,i}(\bar q_{\mathrm r})$ corresponds to some element in the main diagonal of  $-\mathrm{diag}( \vert  \bar q_{\mathrm r} \vert )$ or  $-\mathrm{diag}(\tilde{ \mathcal{T}} \vert \bar q_{\mathrm{r}}\vert  )$, which are clearly negative. Condition (ii) can be established using similar arguments. Indeed, simply note that each off-diagonal element $\tilde{\mathcal{A}}_{i,j}(\bar{q}_\mathrm{r})$, $i\neq j$, corresponds to either one of the components of $\text{diag}(\vert \bar{q}_\mathrm{r}\vert)\tilde{\mathcal{S}}^\top$ or $\tilde{\mathcal{T}}\text{diag}(\vert \bar{q}_\mathrm{r}\vert)$, which are all non negative.} \revtwoarxiv{To verify conditions (iii) and (iv) let us compute on the one hand the following:
\begin{align*}
\tilde{\mathcal{A}}(\bar q_{\mathrm r})\boldsymbol{1}  & = \begin{bmatrix}
-\mathrm{diag} ( \vert \bar{q}_\mathrm{r} \vert ) \boldsymbol{1}+ \mathrm{diag}( \vert \bar{q}_\mathrm{r} \vert ) \tilde{\mathcal{S}}^\top \boldsymbol{1}\\
\tilde{\mathcal{T}} \mathrm{diag}( \vert \bar q_\mathrm{r} \vert ) \boldsymbol{1}- \mathrm{diag}(\tilde{\mathcal{T}} \vert \bar q_\mathrm{r} \vert ) \boldsymbol{1}
\end{bmatrix} \\
& = \begin{bmatrix}
- \vert \bar{q}_\mathrm{r} \vert + \vert \bar{q}_\mathrm{r} \vert\\
\tilde{\mathcal{T}}\vert \bar{q}_\mathrm{r} \vert -\tilde{\mathcal{T}}\vert \bar{q}_\mathrm{r} \vert 
\end{bmatrix}  = \boldsymbol{0},
\end{align*}
where we have used the fact that $\tilde{\mathcal{S}}^\top \boldsymbol{1}=\boldsymbol{1}$ (see \eqref{eq:incidence_target}). On the other hand, it can  be verified that
\begin{align*}
& \boldsymbol{1}^\top \mathcal{A}(\bar{q}_\mathrm{r})\\
 & = \begin{bmatrix}
-\vert \bar{q}_\mathrm{r} \vert ^\top + \vert \bar{q}_\mathrm{r} \vert ^\top  & \vdots & \vert \bar{q}_\mathrm{r} \vert^\top \tilde{\mathcal{S}}^\top - \vert \bar{q}_\mathrm{r} \vert^\top \tilde{\mathcal{T}}^\top
\end{bmatrix}\\
& = \begin{bmatrix}
\boldsymbol{0}^\top & \vdots & -\left( \left(\tilde{\mathcal{T}}-\tilde{\mathcal{S}} \right)\vert  \bar{q}_\mathrm{r}\vert \right)^\top
\end{bmatrix} = \begin{bmatrix}
\boldsymbol{0}^\top & \vdots & -\left(\tilde{\mathcal{B}}\vert  \bar{q}_\mathrm{r}\vert \right)^\top
\end{bmatrix}.
\end{align*}
In this case we have used the identity $\boldsymbol{1}^\top \tilde{\mathcal{T}}=\boldsymbol{1}^\top$.  To see that $\boldsymbol{1}^\top \tilde{\mathcal{A}}(\bar{q}_\mathrm{r})=\boldsymbol{0}^\top$, we note that, at $\bar{q}_\mathrm{r}$, the node volume dynamics satisfies $\dot{\tilde{V}}_\mathrm{N}=\tilde{\mathcal{B}}_0 \bar{q}_\mathrm{r}=\tilde{\mathcal{B}} \vert \bar{q}_\mathrm{r} \vert = \boldsymbol{0} $.}

In view of the above,   $\tilde{\mathcal{A}}(\bar q_{\mathrm r})$ is a KCM, and hence  negative semi definite \cite[Lemma~7]{hangos_thermo_99}; \revone{{\em c.f.}, \cite[Theorem~1]{rein_rom_2021}. Since $A_\mathrm{th}(\bar{q}_\mathrm{r})$ is the Schur complement of an appropriate sub-matrix of $\tilde{\mathcal{A}}(\bar{q}_\mathrm{r})$, then it  is also negative semi definite.}\footnote{\revtwoarxiv{See Appendix~\ref{app:further_details_proof_lemma_calAnegdef} for the case in which some components of $\bar{q}_\mathrm{r}$ are zero.}}~$\hfill\blacksquare$

\end{pf}

The main result of this subsection is the following:
\begin{theorem}\label{theorem:shifted_pass_TDYN}
The DH system's thermal dynamics~\eqref{eq:TDYN_redu} are  shifted passive with storage function $\mathcal{H}_{\mathrm{th}}(T_{\mathrm{th}})=\frac{1}{2}(T_{\mathrm{th}}-\bar{T}_{\mathrm{th}})^\top  \mathrm{diag}( \bar{V}_\mathrm{th} ) (T_{\mathrm{th}}-\bar{T}_{\mathrm{th}})$ 
and shifted passive output $y_{\mathrm{th}}-\bar{y}_{\mathrm{th}} = \revone{B_{\mathrm{pr}}}^\top  (T_{\mathrm{th}}-\bar{T}_{\mathrm{th}})$,  under the  assumption that $q_{\mathrm r}=\bar q_{\mathrm r}$ \revone{and $P_{\mathrm{c}}$} are constant  all the time.
\end{theorem}

\begin{pf}
\revtwo{Let us assume that the consumers' power $P_{\mathrm{c}}$ is constant.} Let $(\bar q_{\mathrm r}, \bar T_{\mathrm{th}}, \bar P_{\mathrm{pr}})$ \revtwo{be} an associated equilibrium triple for   \eqref{eq:TDYN_redu}, {\em i.e.}, the identity $\mathrm{diag}( \bar{V}_\mathrm{th} ) \dot{\bar{T}}_{\mathrm{th}}=\boldsymbol{0}= A_{\mathrm{th}}(\bar q_{\mathrm r})\bar T_{\mathrm{th}}+\revone{B_{\mathrm{pr}} \bar P_{\mathrm{pr}} - B_\mathrm{c} P_\mathrm{c}}$  holds,  where  $\mathrm{diag}(  \bar{V}_\mathrm{th})$ is constant  and positive definite (see Remark~\ref{rem:diff_claudios_etc}.(i)).  Since  $q_{\mathrm r}=\bar q_{\mathrm r}$ all the time, then \eqref{eq:TDYN_redu} is equivalent to 
\begin{equation}\label{eq:shifted_dyn_Tdyn}
\mathrm{diag}( \bar{V}_{\mathrm{th}} ) \dot{T}_{\mathrm{th}} = A_{\mathrm{th}}(\bar q_{\mathrm r})\left(T_{\mathrm{th}}-\bar{T}_{\mathrm{th}} \right)+\revone{B_{\mathrm{pr}} \left( P_{\mathrm{pr}}- \bar{P}_{\mathrm{pr}} \right).}
\end{equation}
Let us consider the mapping $\mathcal{H}_{\mathrm{th}}$ as a candidate storage function. Clearly $\mathcal{H}_{\mathrm{th}}$ is continuously differentiable and  bounded from below by zero. Moreover, along the solutions of  \eqref{eq:shifted_dyn_Tdyn}, its derivative can be shown to satisfy $\dot{\mathcal{H}}_{\mathrm{th}}(T_{\mathrm{th}}) \leq (y_{\mathrm{th}}-\bar y_{\mathrm{th}})^\top (\revone{P_{\mathrm{pr}}-\bar P_{\mathrm{pr}}})$, where we have used the fact that matrix $A_{\mathrm{th}}(\bar q_{\mathrm r})$ is negative semi definite (see Lemma~\ref{lem:calA_neg_sem}). This concludes the proof. $\hfill\blacksquare$
\end{pf}

\begin{remark}\label{rem:shifted_passivity}
\smallskip

\begin{enumerate}[label=(\roman*), wide, labelwidth=!, labelindent=0pt]

\item  In view of the potential time-scale separation between hydraulic (fast) and thermal (slow) dynamics,   
 and under certain conditions (see \cite[Theorem~1]{hildeberto_slowfast_17} and \cite[Theorem 11.4]{khalil_book_nls_02}),   singular perturbation theory \cite{kokotovic_spm_86} is useful to describe the qualitative behavior of the overall slow-fast system  by separately analyzing  each subsystem. This is the main reason why we directly analyze the slow subsystem assuming that $q_{\mathrm r}=\bar q_{\mathrm r}$ all the time. 

\smallskip

\item \revone{The storage function  of the flow dynamics~\eqref{eq:flow_dynamics_qf} corresponds to the total kinetic energy of the stream through each pipe in the system, shifted with respect to an equilibrium value. Also, the passive output is a vector stacking the flows through the DH system's chords.}

\smallskip

\item The proposed storage function $\mathcal{H}_{\mathrm{th}}$ in Theorem~\ref{theorem:shifted_pass_TDYN} is the DH system's shifted ectropy (see \cite{haddad_book_new}),  which is  quadratic in the total energy of the system \cite[Chapter 3]{haddad_book_new}; see also \cite{dong_passivity_19} \revone{and \cite[Theorem~3.1]{hauschild_ph_20}}. \revone{To see this, we recall that potential, kinetic and flow energies were neglected to establish the DH system's thermal dynamics and that we assumed that  the internal energy of any element $i\in\mathcal{E}\cup \mathcal{N}$ is linear with respect to its temperature. Also, we recall that we took $\rho=c_\mathrm{s.h.}=1$ to simplify notation.} Moreover, considering the definition of the matrix $B_{\mathrm{pr}}$ in Corollary~\ref{corollary: thermal_state_dyn}, then   the  passive output $y_{\mathrm{th}}=B_{\mathrm{pr}}^\top T_{\mathrm{th}}$ corresponds to a vector  stacking the (shifted) \revone{internal energies} of the producers' heat exchangers (see \eqref{eq:app_pipe_prod_cons}) to which the power $P_{{\mathrm{pr}},i}$ is delivered.

\smallskip

\item \revone{Based on the results of \cite{bayu_scl_07,DePersis2011}, \revtwoarxiv{in Appendix~\ref{app:usefulness_passivity}}  we discuss the usefulness of the shifted passivity properties identified in Theorem~\ref{theorem:shifted_passivity} and \ref{theorem:shifted_pass_TDYN} for the design of decentralized, passivity-based PI controllers to regulate the passive outputs towards desired, constant setpoints.  The conditions to guarantee  closed-loop asymptotic stability, which rely on verifying a detectability condition for the passive output  (see \cite{arjan_l2gain_book}),  are mentioned as well. Moreover, the results are supported by numerical simulations on the model \eqref{eq:TDYN_redu} \revtwoarxiv{in Section~\ref{sec:simulations}}. System pressures and storage volume regulation are addressed in  \cite{DePersis2014} and \cite{jmnj_css_21}, respectively.

\smallskip

\item \revtwo{As we specified at the beginning of this section, we would like to  remark that Lemma~\ref{lem:calA_neg_sem} and Theorem~\ref{theorem:shifted_pass_TDYN} hold also for \eqref{eq:dyntemp_nonred}; notice that $\mathcal{A}(\bar{q}_\mathrm{E})$ has the same structure and properties of $\tilde{\mathcal{A}}(\bar{q}_\mathrm{E})$.}

}

\end{enumerate}

\end{remark}

\section{Simulations}\label{sec:simulations}

The derived model and its properties are illustrated in this section  through a numerical simulation. We consider a DH system comprising three heat producers ($n_\mathrm{pr}=3$) and nine consumers which are interconnected by a meshed distribution network as sketched in Fig.~\ref{fig:dh_meshed}.
The maximum thermal power injection for the producers is set as $P_{\mathrm{pr}}^{\max}=[25, 15, 35]~\mathrm{MW}$, whereas the demand from the consumers is fixed at $P_{\mathrm{c}}=[5,6,9,7,6,4,10,7,5]~\mathrm{MW}$. The designed difference between supply and return temperatures is $30^\circ\mathrm{C}$. Moreover, every storage tank is assumed to have a total capacity of $1000~\mathrm{m}^3$.
%
%
The density and  specific heat are respectively taken as $\rho=975~\mathrm{kg}/\mathrm{m}^3$ and $c_{\mathrm{s.h.}}=4190~\mathrm{J}/\mathrm{kg}^\circ\mathrm{C}$. Also, the pipe roughness and  water viscosity are $\epsilon=0.5\times 10^{-3}\mathrm{m}$ and $\eta=3\times 10^{-4}\mathrm{kg}/\mathrm{m}\mathrm{s}$, respectively.\footnote{The  selection of the system parameters, including pipe diameters and lengths, was based on the design guidelines  and the case studies reported in \cite{palsson_book_99,
wang_meshed_17,
Scholten_tcst_2015,
hassine_pipe_13,
dominkovic_cluster_20,
Gabrielaitiene2007,
grosswindhager_delay_11,
torres_hamiltonian_19}.}
%
%

We consider two time intervals given by $\mathcal{I}_1:0\leq t< 5.4\times 10^4~\mathrm{s}$ and $\mathcal{I}_2:5.4\times 10^4\leq t \leq 10.8\times 10^4~\mathrm{s}$, where the power demand from the consumers is kept constant for all time, but  the setpoints for the producers' temperatures is piecewise constant with different values in each interval. 
The flow dynamics (see Theorem~\ref{theorem:overall_flows})  are in closed-loop with the controller \eqref{eq:pi_chords}, for which we take $k^\mathrm{p}_{\mathrm{f},i}=k^\mathrm{I}_{\mathrm{f},i}=10^{7}$, $i=1,\dots,n_\mathrm{f}$. 
Moreover, the system is assumed to have a single boost pump which is placed in the distribution side of the storage tank associated to Producer~2, and is considered to maintain a constant pressure  difference of $8.28~\mathrm{bar}$ between its terminals. On the other hand,  the thermal dynamics (see Corollary~\ref{corollary: thermal_state_dyn}) are  in closed-loop with the controller \eqref{eq:pi_Tth}, for which we select the gains $k^\mathrm{p}_{\mathrm{th},i}=k^{\mathrm{I}}_{\mathrm{th},i}={\color{black}\tfrac{1}{\rho c_\mathrm{s.h.}}}10^{10}$, $i=1,\dots,n_\mathrm{pr}$. The desired supply temperature $\bar T_{\mathrm{pr}}$ is set as $85\, 1_{n_\mathrm{pr}}(^\circ\mathrm{C})$ in the interval $\mathcal{I}_1$ and as $\left[85~87~83 \right]^\top(^\circ\mathrm{C})$ for $\mathcal{I}_2$.\footnote{The control gains pairs $(k^\mathrm{p}_{\mathrm{f},i},k^\mathrm{I}_{\mathrm{f},i})$ and $(k^\mathrm{p}_{\mathrm{th},i},k^{\mathrm{I}}_{\mathrm{th},i})$   have been chosen via trial-and-error so that a fair trade-off between settling time and overshoot was attained.}

\begin{figure}
\begin{subfigure}{\linewidth}
\includegraphics[width=\textwidth]{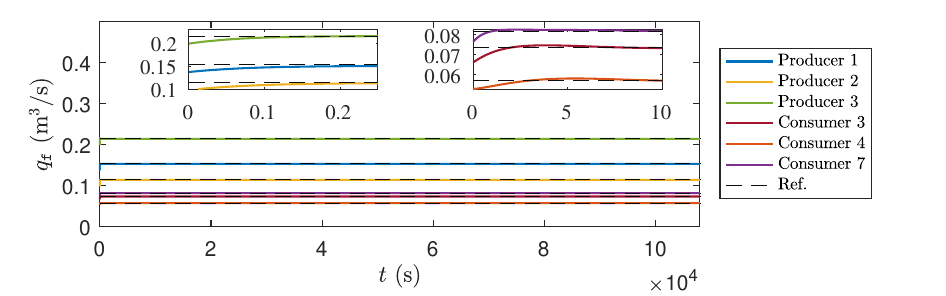}
\end{subfigure}
\begin{subfigure}{\linewidth}
\includegraphics[width=\textwidth]{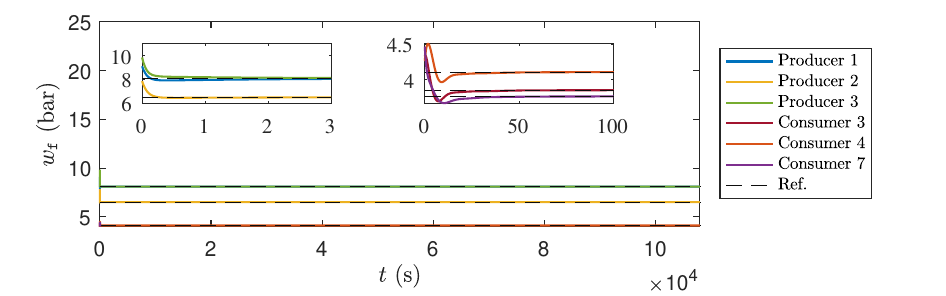}
\end{subfigure}
\begin{subfigure}{\linewidth}
\includegraphics[width=\textwidth]{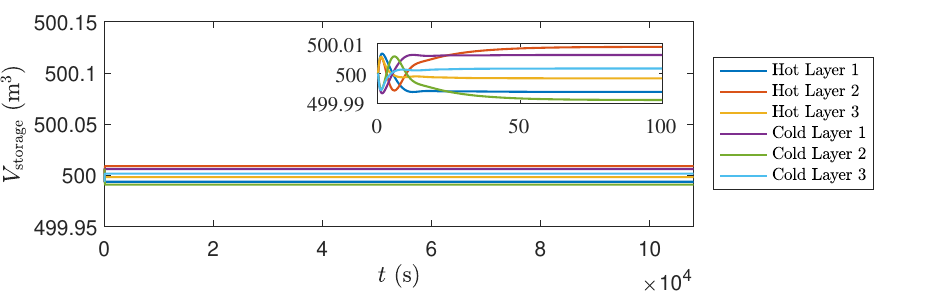}
\end{subfigure}
\caption{Simulation results hydraulic dynamics.}
\label{fig:sim_res_hydr}
\end{figure}

\begin{figure}
\begin{subfigure}{\linewidth}
\includegraphics[width=\textwidth]{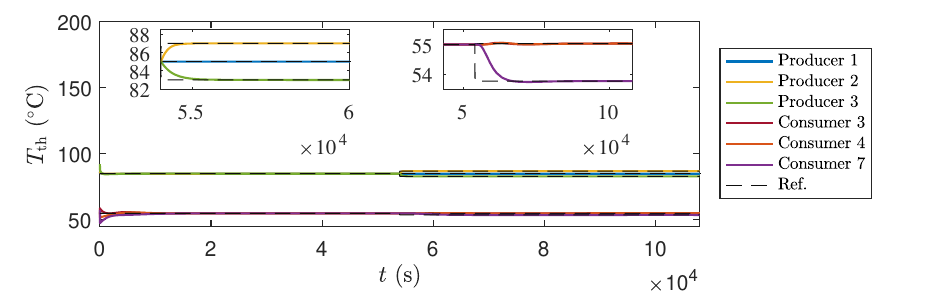}
\end{subfigure}
\begin{subfigure}{\linewidth}
\includegraphics[width=\textwidth]{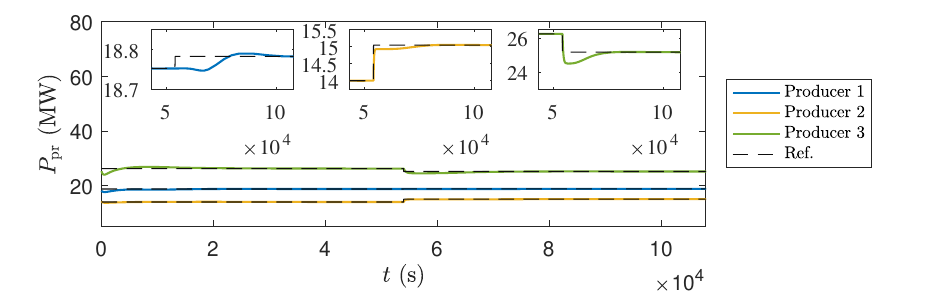}
\end{subfigure}
\caption{Simulation results thermal dynamics.}
\label{fig:sim_res_thermal}
\end{figure}

The description of the simulations  results shown in Figs.~\ref{fig:sim_res_hydr} and  \ref{fig:sim_res_thermal} is as follows. The initial conditions of the overall closed-loop DH dynamics were randomly selected near the equilibrium, with the exception of the  storage tanks hot and cold layers' volumes which, for simplicity, were all initialized at $500~\mathrm{m}^3$.
It can be seen in Fig.~\ref{fig:sim_res_hydr} that all plotted components of the flow vector $q_\mathrm{f}$ asymptotically converge to (prespecified) reference values. Only a selected number of components are plotted to enhance readability. We note also that hot and cold layers  of the storage tanks are  bounded around the initial condition and converge to some stationary state (we recall that we have not closed any loop around the volume of the tanks).

In Fig.~\ref{fig:sim_res_thermal} we see the supply temperatures of all producers reaching their desired (piecewise constant) reference values. The evolution of the power injected by the producers is also shown, where convergence to their respective equilibrium values is evident. We note that in the interval $\mathcal{I}_2$ there is a change in the injected power proportions. This is consistent with the modified  setpoints of the producers' temperatures. For the case of the  consumers,  we observe that all plotted temperatures also attain a constant equilibrium in each interval, but at a much lower rate with respect to the producers, which is an expected behavior.  

\section{Conclusions}\label{sec:conclusions}

In this  paper we have addressed the modeling of a district heating system containing multiple heat producers, storage devices and consumers. All these elements are assumed to be interconnected  through a common distribution network of meshed topology. The derived  model is highly nonlinear, uncertain and explicitly considers the temperature dynamics of the pipes in the distribution network. We have analyzed the conditions under which this model is shifted passive,    opening the possibility for the design of passivity-based controllers of simple structure that additionally offer closed-loop stability guarantees. 
Future research directions include: (i) the relaxation of some of our modeling assumptions, {\em e.g.}, the consideration that there are no stand-alone storage tanks; (ii) the consideration of more general heat demand profiles from consumers ({\em c.f.}, \cite{Scholten_tcst_2015}); and (iii) the study of the usefulness of the established  passivity properties for  the design of distributed or possibly hierarchical passivity-based control  schemes such that the flows and temperatures are regulated towards optimal setpoints (see \cite[Section~6.3]{vandermeulen_control_review_18}).

\appendix

\section{Symbols Description}\label{app:summary_table}
\medskip


{\footnotesize

\begin{tabular}{ll}
$\mathcal{G}$, $\mathcal{N}$, $\mathcal{E}$ & DH system's graph, nodes and edges\\
$\mathcal{B}_0$ & Arbitrary incidence matrix of $\mathcal{G}$\\
$\mathcal{B}$ & Incidence matrix of $\mathcal{G}$ where the orientation\\ 
~ & of the edges  matches the streams' directions\\
$G$ & reduced matrix obtained by viewing each tank\\
~ &  as a node\\
$\mathcal{L}_i$ & $i$th fundamental  loop of $G$\\
$F$ & Fundamental loop matrix of $G$\\
$P_{\mathrm{pr},i}$ & Power injection by $i$th producer, W\\
$P_{\mathrm{c},i}$ & Power extraction by $i$th consumer, W\\
$q_{\mathrm{E},i}$ & flow rate  through edge $i\in \mathcal{E}$, $\mathrm{m}^3/s$\\
$p_{i}$ & Pressure of the $i\in \mathcal{N}\cup \mathcal{E}$, Pa\\
$V_{i}$ & Volume of  $i\in \mathcal{N}\cup \mathcal{E}$, $\mathrm{m}^3$\\
$T_{i}$ & Temperature of $i\in \mathcal{N}\cup \mathcal{E}$, $~^\circ \mathrm{C}$\\
$f_{\mathrm{E},i}$ & Pressure drop due to viscous friction of the\\
~ &  $i$th pipe, Pa\\
$w_{\mathrm{E},i}$ & Pressure by $i$th pump, Pa\\
$\ell_i$ & Length of the $i$th pipe, m\\
$A_{i}$ & Cross-section area of the $i$th pipe, $\mathrm{m}^2$\\
$K_{\mathrm{E},i}$ & Friction factor associated to $f_{\mathrm{E},i}$,\\
$\rho$ & Density of water, $\mathrm{kg}/\mathrm{m}^3$\\
$c_{\mathrm{s.h.}}$ & Specific heat of water, $\mathrm{J}/\mathrm{kg}~^\circ \mathrm{C}$\\
$(\cdot)_{\mathrm{pr},i}$ & quantity associated to the $i$th producer\\
$(\cdot)_{\mathrm{f},i}$ & quantity associated to the $i$th chord\\
$(\cdot)^\mathrm{in}_i$ & quantity associated to an inlet\\
$(\cdot)^\mathrm{out}_i$ & quantity associated to an outlet
\end{tabular}


}

\section{Further details of the proof of Proposition~\ref{prop:hydraulic_DAE}}\label{app:further_details_prop_flow_DAE}

By recalling that water density $\rho$ is constant by assumption, then the differential  forms of the balance of mass and momentum at any pipe can be written as follows (see \cite{hauschild_ph_20}):
\begin{subequations}\label{eq:PDE_flow_pipe}
\begin{align}
0  & = \partial_x v_{\mathrm{E},i}(t,x)\\
0 & = \rho \partial_t v_{\mathrm{E},i}(t,x)+\partial_x p_{\mathrm{E},i}(t,x)+f_{\mathrm{E},i}(t,x),
\end{align}
\end{subequations}
where $v_{\mathrm{E},i}(t,x)$ and $p_{\mathrm{E},i}(t,x)$ are the velocity and the pressure of the stream along the  axis of the pipe ($0 \leq x \leq \ell_{\mathrm{E},i}$). Let us assign  $q_{\mathrm{E},i}(t):=A_{\mathrm{E},i}v_{\mathrm{E},i}(t,x)$,  $p_{\mathrm{E},i}^\mathrm{in}:=p_{\mathrm{E},i}(t,0)$, $p_{\mathrm{E},i}^\mathrm{out}:=p_{\mathrm{E},i}(t,\ell_{\mathrm{E},i})$ and  consider
$$
f_{\mathrm{E},i}(t,x)=\frac{\theta_{\mathrm{E},i} \rho \ell_{\mathrm{E},i}}{2 D_{\mathrm{E},i}A_{\mathrm{E},i}^2}\vert q_{\mathrm{E},i}\vert q_{\mathrm{E},i}
$$
with the assumption that the Darcy friction factor $\theta_{\mathrm{E},i}$  is uniform along the pipe's axis and constant in time. Then,  each equation in \eqref{eq:PDE_flow_pipe}  can be integrated with respect to $x$ to give  \eqref{eq_rev30:press_drop}.

\section{Further details of the proof of Theorem~\ref{theorem:overall_flows}}\label{app:further_details_proof_theorem_flows}
To see that  $\mathcal{J}_{\mathrm{f}}$  is positive definite consider the following (the procedure follows the developments reported in \cite{DePersis2011}). Recall that $J_{\mathrm{E},i}>0$ for any  $i\in \mathcal{C}$ (see the selection of the set of chords $\mathcal{C}$ in page 6, right column), whereas  $J_{\mathrm{E},i}\geq 0$ for each $i\in \mathcal{E}\setminus \mathcal{C}$. We saw that  $\mathcal{E}$ can be ordered as  $\mathcal{C}=\mathcal{C}\cup \mathcal{W}_\mathrm{f}\cup \mathcal{E}''$ without loss of generality. Then,  $\mathrm{diag}(J_{\mathrm{E},i})_{i\in \mathcal{E}}$ can be written as follows:
$$
\mathrm{diag}(J_{\mathrm{E},i})_{i\in \mathcal{E}}=\mathrm{diag}\left(J_{\mathrm{f}},~J_{\mathrm g} \right),
$$
where $J_{\mathrm{f}}=\mathrm{diag}(J_{\mathrm{E},i})_{i\in \mathcal{C}}$,  is positive definite and $J_{\mathrm g}=\mathrm{diag}( J_{\mathrm{E},i})_{i\in \mathcal{W}_\mathrm{f}\cup \mathcal{E}''}$,  is positive semi definite. Considering \eqref{eq:split_fund_loop_mat}, it follows that
\begin{align}\label{eq:detail_JF>0}
\mathcal{J}_{\mathrm{f}} & =  F \mathrm{diag}(J_{\mathrm{E},i})_{i\in \mathcal{E}} F^\top \nonumber \\
&  = \left[ I_{n_{\mathrm{f}}}~H^\top  \right] \begin{bmatrix}
J_{\mathrm{f}} & \boldsymbol{0}\\
\boldsymbol{0} & J_{\mathrm g}
\end{bmatrix} \begin{bmatrix}
I_{n_{\mathrm{f}}}\\
H
\end{bmatrix} \nonumber \\
& = J_{\mathrm{f}} + HJ_{\mathrm g} H^\top,
\end{align}
where $H=\begin{bmatrix} I_{\mathrm{n}_\mathrm{f}} & B_{\mathrm{b}} \end{bmatrix}$. From \eqref{eq:detail_JF>0} it can be followed that   $\mathcal{J}_{\mathrm{f}}$ is indeed positive definite.

\section{A corollary to Theorem~\ref{theorem:overall_flows}}\label{app:a_decoupling}

\begin{corollary}\label{cor:flow_dyn_decomp}
Let  $q_\mathrm{ch}$ and $q_\mathrm{pr}$ denote the flow vectors of the chords  $\mathcal{P}_\mathrm{c}\cup \mathcal{P}_\mathrm{d}\cup \mathcal{P}_\mathrm{ST} $ and $ \mathcal{P}_\mathrm{pr}$, respectively.   Then, there exist matrices $J_\mathrm{ch}$,  $J_\mathrm{pr}$ and mappings  $f_\mathrm{ch}$,  $f_\mathrm{pr}$ such that  \eqref{eq:flow_dynamics_qf} is equivalent to
\begin{subequations}
\begin{align}
J_\mathrm{ch}\dot{q}_\mathrm{ch} & = -f_\mathrm{ch}(q_\mathrm{ch})+w_\mathrm{ch}+\hat{B}_\mathrm{ch}\hat{w}_\mathrm{ch}\\
J_\mathrm{pr} \dot{q}_\mathrm{pr} & =-f_\mathrm{pr}(q_\mathrm{pr})+w_\mathrm{pr}+\hat{B}_\mathrm{pr}\hat{w}_\mathrm{pr},
\end{align}
\end{subequations}
where $w_{\mathrm{ch},i}$ and $w_{\mathrm{pr},j}$ are the pressures produced by pumps adjacent to $i\in \mathcal{P}_\mathrm{c}\cup \mathcal{P}_\mathrm{d}\cup \mathcal{P}_\mathrm{ST} $ and $j\in \mathcal{P}_\mathrm{pr}$, respectively and $\hat{B}_\mathrm{ch}\hat{w}_\mathrm{ch}$ and  $\hat{B}_\mathrm{pr}\hat{w}_\mathrm{pr}$, with $(\hat{B}_\mathrm{ch})_{\alpha,\beta}, (\hat{B}_\mathrm{pr})_{\alpha,\beta}\in \{-1,0,1 \}$,  codify the effect of booster pumps on $q_\mathrm{ch}$ and $q_\mathrm{pr}$, respectively. Moreover, $J_\mathrm{ch}$ is symmetric, positive definite and $J_\mathrm{pr}$ is diagonal, positive definite. In addition, the Jacobian matrix of $f_\mathrm{pr}$ is diagonal.
\end{corollary}

\begin{pf}
{For ease of notation, in this proof we define  $\mathcal{P}_\mathrm{ch}:=\mathcal{P}_\mathrm{c}\cup \mathcal{P}_\mathrm{d}\cup \mathcal{P}_\mathrm{ST}$. Also, we assume out of simplicity that for each storage tank the only pipe, valve and pump in its producer side are those of the associated producer  (see Fig.~\ref{fig:prod_stor}). Then,  the series-connected pump, pipe and valve of any   given  producer form a fundamental loop (viewing each tank as a node of $G$). Notably, this fundamental loop excludes any other edge of $\mathcal{E}$. This also  implies  that fundamental loops associated to any chord in $\mathcal{P}_\mathrm{ch}$ excludes all edges associated to producers. More precisely, if we let $\mathcal{H}_\mathrm{ch}$ and $\mathcal{H}_\mathrm{pr}$ denote the set of all edges in all the fundamental loops associated to $\mathcal{P}_\mathrm{ch}$ and $\mathcal{P}_\mathrm{pr}$, respectively, then $\mathcal{E}=\mathcal{H}_\mathrm{ch}\cup \mathcal{H}_\mathrm{pr}$ and $\mathcal{H}_\mathrm{ch}\cap \mathcal{H}_\mathrm{pr} =\emptyset$.

Considering  the above,  it is possible to follow the arguments in the proof of Theorem~\ref{theorem:overall_flows},  to order $\mathcal{E}$  and $\mathcal{C}=\mathcal{P}_\mathrm{ch}\cup \mathcal{P}_\mathrm{pr}$ and partition the   fundamental loop matrix $F$ as
\begin{equation}\label{eq:fund_loop_mat_chords_prod}
F=\begin{bmatrix}
I_{n_\mathrm{ch}} & \mathbf{0} & H_{\mathrm{ch}} &  \mathbf{0}\\
\mathbf{0} & I_{n_\mathrm{pr}} & \mathbf{0} & H_{\mathrm{pr}}
\end{bmatrix},
\end{equation}
where $n_\mathrm{ch}$ and $n_\mathrm{pr}$ are the cardinalities of $\mathcal{P}_\mathrm{ch}$ and $\mathcal{P}_\mathrm{pr}$, respectively, and
\begin{subequations}\label{eq:decomp_H_ch_H_pr}
\begin{align}
H_\mathrm{ch} & = \begin{bmatrix}I_{n_\mathrm{ch}} &\hat{B}_\mathrm{ch}\end{bmatrix},\\
H_{\mathrm{pr}} & = \begin{bmatrix}
I_{n_\mathrm{pr}} & \hat{B}_\mathrm{pr}
\end{bmatrix},
\end{align}
\end{subequations}
where the first $n_\mathrm{ch}$ ($n_\mathrm{pr}$) columns of $H_\mathrm{ch}$ ($H_\mathrm{pr}$) would be associated to the pump, with pressure $w_{\mathrm{ch},i}$ ($w_{\mathrm{pr},i}$) adjacent to $i\in \mathcal{P}_\mathrm{ch}$ ($i\in \mathcal{P}_\mathrm{pr}$). We note  that by considering the simple topology of  producers and their associated fundamental loops, $\hat{B}_\mathrm{pr}$ is a diagonal matrix.

Now, in view of \eqref{eq:fund_loop_mat_chords_prod},  we get by direct computations   that $\mathcal{J}_\mathrm{f}$, as defined in Theorem~\ref{theorem:overall_flows}, can be written as follows:
\begin{align*}
\mathcal{J}_\mathrm{f} & = F\mathrm{diag}(J_\mathrm{E})F^\top=\mathrm{diag}(J_\mathrm{ch}, J_\mathrm{pr}),
\end{align*}
where 
\begin{align*}
J_\mathrm{ch} & = \mathrm{diag}(J_{\mathrm{E},i})_{i\in \mathcal{P}_\mathrm{ch}}+H_\mathrm{ch} \mathrm{diag}(J_{\mathrm{E},i})_{i\in \mathcal{H}_\mathrm{ch}}H_\mathrm{ch}^\top,\\
J_\mathrm{pr} & = \mathrm{diag}(J_\mathrm{E})_{i\in \mathcal{P}_\mathrm{pr}}+ H_\mathrm{pr} \mathrm{diag}(J_{\mathrm{E},i})_{i\in \mathcal{H}_\mathrm{pr}}H_\mathrm{pr}^\top.
\end{align*}
This establishes the stated properties for $J_\mathrm{ch}$ and $J_\mathrm{pr}$.

Now we write the explicit forms of $f_\mathrm{ch}$ and $f_\mathrm{pr}$ from $f_\mathrm{f}$. First, let us compute the following:
\begin{align}\label{eq:explicit_form_f_pr_f_ch_0}
&f_{\mathrm{f}}(q_\mathrm{f})  = Ff_{\mathrm{E}}(q_\mathrm{E})\vert_{q_\mathrm{E}=F^\top q_\mathrm{f}} \nonumber \\
& =\resizebox{0.4\textwidth}{!}{$ \left. \begin{bmatrix}\mathrm{col}(f_{\mathrm{E},i}(q_{\mathrm{E},i}))_{i\in \mathcal{P}_\mathrm{ch}} + H_\mathrm{ch}\mathrm{col}(f_{\mathrm{E},i}(q_{\mathrm{E},i}))_{i\in \mathcal{H}_\mathrm{ch}}\\
\mathrm{col}(f_{\mathrm{E},i}(q_{\mathrm{E},i}))_{i\in \mathcal{P}_\mathrm{pr}} + H_\mathrm{pr}\mathrm{col}(f_{\mathrm{E},i}(q_{\mathrm{E},i}))_{i\in \mathcal{H}_\mathrm{pr}}
\end{bmatrix} \right\vert_{q_{\mathrm{E}}=F^\top q_\mathrm{f}}$},
\end{align}
where for clarity  we use, particularly in this proof, $\mathrm{col}(x_i)$ to represent a column vector. 
Now, taking into  consideration that $q_\mathrm{f}$ can be written as $q_\mathrm{f}=[q_\mathrm{ch}^\top ~~ q_\mathrm{pr}^\top]^\top$ (see \eqref{eq:fund_loop_mat_chords_prod}), then $q_\mathrm{E}$ is equivalent to
\begin{align*}
q_{\mathrm{E}}=F^\top q_\mathrm{f}=\begin{bmatrix}q_\mathrm{ch}\\
q_\mathrm{pr}\\
H_\mathrm{ch} ^\top q_\mathrm{ch}\\
H_\mathrm{pr}^\top q_\mathrm{pr}
\end{bmatrix}.
\end{align*}
It follows that $f_\mathrm{f}(q_\mathrm{f})$ in \eqref{eq:explicit_form_f_pr_f_ch_0} can be partitioned as
\begin{align}\label{eq:explicit_form_f_pr_f_ch_1}
 f_\mathrm{f}(q_\mathrm{f})= \begin{bmatrix}
f_\mathrm{ch}(q_\mathrm{ch})\\
f_\mathrm{pr}(q_\mathrm{pr})
\end{bmatrix},
\end{align}
where, with some  abuse of notation,
\begin{subequations}\label{eq:explicit_form_f_pr_f_ch_2}
\begin{align}
f_\mathrm{ch}(q_\mathrm{ch}) & = \mathrm{col}(f_{\mathrm{E},i})_{i\in \mathcal{P}_\mathrm{ch}}\circ \mathrm{id}(q_\mathrm{ch}) \nonumber \\
&~~~~~+ H_\mathrm{ch} \mathrm{col}(f_{\mathrm{E},i})_{i\in \mathcal{H}_\mathrm{ch}}\circ \mathrm{id}(H_\mathrm{ch}^\top q_\mathrm{ch})\\
f_\mathrm{pr}(q_\mathrm{pr}) & = \mathrm{col}(f_{\mathrm{E},i})_{i\in \mathcal{P}_\mathrm{pr}}\circ \mathrm{id}(q_\mathrm{pr}) \nonumber \\
&~~~~~+ H_\mathrm{pr} \mathrm{col}(f_{\mathrm{E},i})_{i\in \mathcal{H}_\mathrm{pr}}\circ \mathrm{id}(H_\mathrm{pr}^\top q_\mathrm{pr}).
\end{align}
\end{subequations}
Here $\mathrm{id}(\cdot)$ represents the identity function. From the above  definition of $f_\mathrm{pr}$ and considering the structure of $H_\mathrm{pr}$ in \eqref{eq:decomp_H_ch_H_pr}, it can be concluded also  that the Jacobian matrix of $f_\mathrm{pr}$ is diagonal. 

The proof that 
\begin{align*}
Fw_\mathrm{E}=\begin{bmatrix}
w_\mathrm{ch}+\hat{B}_\mathrm{ch}\hat{w}_\mathrm{ch}\\
w_\mathrm{pr}+\hat{B}_\mathrm{pr}\hat{w}_\mathrm{pr}
\end{bmatrix}
\end{align*}
is straightforward considering \eqref{eq:fund_loop_mat_chords_prod} and \eqref{eq:decomp_H_ch_H_pr}.  $\hfill\blacksquare$


}
\end{pf}

\section{The system \eqref{eq:dyntemp_nonred} as a semi-explicit DAE of differentiation index one}\label{app:diff_index_1_noncontractedgraph}

Let $\mathcal{E}_0\subset \mathcal{E}$ denote the set of all valves and pumps  in the DH system,  and let $\mathcal{N}_0\subset \mathcal{N}$ denote the set of all simple junctions. Assume that $V_{\mathrm{E},i}=0$ for each $i\in \mathcal{E}_0$ and that $V_{\mathrm{N},j}=0$ for each $j\in \mathcal{N}_0$. Also, let $\mathcal{E}_1=\mathcal{E}\setminus \mathcal{E}_0$ and $\mathcal{N}_1=\mathcal{N}\setminus \mathcal{N}_0$.  Without loss of generality, let us reorder  $\mathcal{E}$ and $\mathcal{N}$ as $\mathcal{E}=\mathcal{E}_1\cup \mathcal{E}_0$ and $\mathcal{N}=\mathcal{N}_1\cup \mathcal{N}_0$, respectively. Let $T_{\mathrm{E}1}=\text{col}(T_{\mathrm{E},i})_{i\in \mathcal{E}_1}$, $T_{\mathrm{E}0}=\text{col}(T_{\mathrm{E},i})_{i\in \mathcal{E}_0}$, $T_{\mathrm{N}1}=\text{col}(T_{\mathrm{N},i})_{i\in \mathcal{N}_1}$ and $T_{\mathrm{N}0}=\text{col}(T_{\mathrm{N},i})_{i\in \mathcal{N}_0}$. Let also analogous definitions hold for $V_{\mathrm{E}1}$, $V_{\mathrm{E}0}$, $V_{\mathrm{N}1}$ and $V_{\mathrm{N}0}$; also for $q_{\mathrm{E}1}$ and $q_{\mathrm{E}0}$. By extending such a reordering to the matrices $\mathcal{T}$ and $\mathcal{S}$ (see \eqref{eq:incidence_target}), we get that they can be partitioned as follows:
\begin{align}
\mathcal{T} & =\begin{bmatrix}
\mathcal{T}_{N1E1} & \mathcal{T}_{N1E0}\\
\mathcal{T}_{N0E1} & \mathcal{T}_{N0E0}
\end{bmatrix},~~~
\mathcal{S} & =\begin{bmatrix}
\mathcal{S}_{N1E1} & \mathcal{S}_{N1E0}\\
\mathcal{S}_{N0E1} & \mathcal{S}_{N0E0}
\end{bmatrix}.
\end{align}
Then, \eqref{eq:dyntemp_nonred} can be written as follows:
\begin{subequations}\label{eq:semi-explicit dae thermal}
\begin{align}
& \text{diag}(V_{\mathrm{E}1})\dot{T}_{\mathrm{E}1}  = -\text{diag}(\vert q_{\mathrm{E}1}\vert)T_{\mathrm{E}1} +\mathbf{P}_{\mathrm{pr}1}-\mathbf{P}_{\mathrm{c}1} \nonumber \\
& ~~ + \text{diag}(\vert q_{\mathrm{E}1}\vert)\mathcal{S}^\top_{E1N1}T_{\mathrm{N}1} + \text{diag}(\vert q_{\mathrm{E}1}\vert)\mathcal{S}^\top_{E1N0}T_{\mathrm{N}0},\\
& 0  =  -\text{diag}(\vert q_{\mathrm{E}0}\vert)T_{\mathrm{E}0} \nonumber \\
& ~~ + \text{diag}(\vert q_{\mathrm{E}0}\vert)\mathcal{S}^\top_{E0N1}T_{\mathrm{N}1} + \text{diag}(\vert q_{\mathrm{E}0}\vert)\mathcal{S}^\top_{E0N0}T_{\mathrm{N}0},\\
& \text{diag}(V_{\mathrm{N}1})\dot{T}_{\mathrm{N}1}  = \mathcal{T}_{N1E1}\text{diag}(\vert q_{\mathrm{E}1} \vert)T_{\mathrm{E}1} \nonumber \\
& ~~ +  \mathcal{T}_{N1E0}\text{diag}(\vert q_{\mathrm{E}0} \vert)T_{\mathrm{E}0} \nonumber \\
& ~~ -\text{diag}(\mathcal{T}_{N1E1}\vert q_{\mathrm{E}1} \vert+ \mathcal{T}_{N1E0}\vert q_{\mathrm{E}0} \vert)T_{\mathrm{N1}}, \\
& 0  = \mathcal{T}_{N0E1}\text{diag}(\vert q_{\mathrm{E}1} \vert)T_{\mathrm{E}1} \nonumber \\
& ~~ +  \mathcal{T}_{N0E0}\text{diag}(\vert q_{\mathrm{E}0} \vert)T_{\mathrm{E}0} \nonumber \\
& ~~ -\text{diag}(\mathcal{T}_{N0E1}\vert q_{\mathrm{E}1} \vert+ \mathcal{T}_{N0E0}\vert q_{\mathrm{E}0} \vert)T_{\mathrm{N0}},
\end{align}
\end{subequations}
where we have used the fact that $V_{\mathrm{E}0}$ and $V_{\mathrm{N}0}$ are  zero vectors, and $\mathbf{P}_{\mathrm{pr}1}=\text{col}(P_{\mathrm{pr},i})_{i\in \mathcal{E}1}$ and $\mathbf{P}_{\mathrm{c}1}=\text{col}(P_{\mathrm{c},i})_{i\in \mathcal{E}1}$. We note that \eqref{eq:semi-explicit dae thermal} is a semi-explicit DAE of the form $\dot{x}=f(x,y)$ and $0=g(x,y)$ (for adequate choices of $x$, $y$, $f$ and $g$). It follows that \eqref{eq:semi-explicit dae thermal} has differentiation index one if the Jacobian of $g$ with respect to $y$ is non-singular. Such a condition is equivalent to the non-singularity of the following matrix:
\begin{equation}\label{eq:difficult nonsingularity}
\begin{bmatrix}
-\text{diag}(\vert q_{\mathrm{E}0}\vert) & \text{diag}(\vert q_{\mathrm{E}0}\vert)\mathcal{S}^\top_{E0N0}\\
\mathcal{T}_{N0E0}\text{diag}(\vert q_{\mathrm{E}0} \vert) & -\text{diag}(\mathcal{T}_{N0E1}\vert q_{\mathrm{E}1} \vert+ \mathcal{T}_{N0E0}\vert q_{\mathrm{E}0} \vert)
\end{bmatrix}.
\end{equation}
Non-singularity of this matrix implies that the algebraic variables  $T_{\mathrm{E}0}$ and $T_{\mathrm{N}0}$ can be explicitly solved from the algebraic constraints in \eqref{eq:semi-explicit dae thermal}, producing in turn a reduced order ODE for the differential variables $T_\mathrm{E1}$ and $T_{\mathrm{N}1}$. The  resulting system is analogous to \eqref{eq:TDYN_redu}, however we did not find general conditions to ensure that \eqref{eq:difficult nonsingularity} is non-singular in general.

\section{Proof of Corollary~\ref{corollary: thermal_state_dyn}}\label{app:proof_corollary_rom}

\begin{pf}
Due to its generality (and since we meet the same set of assumptions), the procedure that led to  \eqref{eq:dyntemp_nonred} can be used to obtain an analogous   model for the temperatures of the devices conforming the  reduced graph $\tilde{\mathcal G}$ obtained by contracting all the edges in $\mathcal{G}$ associated to valves and pumps (whose volumes $V_{\mathrm{E},i}$ we are assuming to be zero). Let $\tilde{\mathcal N}_1=\tilde{\mathcal N}_\mathrm{sh}\cup\tilde{\mathcal N}_\mathrm{sc}$ and let $\tilde{T}_N=\text{col}(\tilde{T}_{N,k})_{j\in \tilde{\mathcal{N}}}$ be split as follows:
\begin{equation}
\tilde{T}_N=\begin{bmatrix}
\tilde{T}_{N1}^\top,& \tilde{T}_{N0}^\top
\end{bmatrix}^\top,
\end{equation}
where $\tilde{T}_{N1}=\text{col}(\tilde{T}_{N,k})_{k\in \tilde{\mathcal N}_1}$ and $\tilde{T}_{N0}=\text{col}(\tilde{T}_{N,k})_{k\in \tilde{\mathcal N}_0}$. Let us consider an analogous ordering (and labeling) for the rows of $\tilde{V}_N=\text{col}(\tilde{V}_{N,j})_{j\in\tilde{\mathcal{N}}}$, as well as the rows  of   the matrices $\tilde{\mathcal T}$ and $\tilde{\mathcal S}$.  Then, the temperature dynamics of the elements of  $\tilde{\mathcal G}$ is analogous to \eqref{eq:dyntemp_nonred} and can thus be written as follows:
\begin{subequations}\label{eq:redu_tem_dyn_comps}
\begin{align}
\text{diag}(\tilde{V}_E)\dot{\tilde{T}}_E & =-\text{diag}(\vert {q}_\mathrm{r}\vert) \tilde{T}_E + \text{diag}(\vert {q}_\mathrm{r} \vert )\tilde{\mathcal{S}}_{N1}^\top \tilde{T}_{N1}\nonumber \\
& ~~~~~ + \text{diag}(\vert {q}_\mathrm{r} \vert )\tilde{\mathcal{S}}_{N0}^\top \tilde{T}_{N0} +\tilde{\mathbf{P}}_{\mathrm{pr}}-\tilde{\mathbf{P}}_{\mathrm{c}}, \label{subeq:edges}\\
\text{diag}(\tilde{V}_{N1})\dot{\tilde{T}}_{N1} & = \tilde{\mathcal{T}}_{N1}^\top \text{diag}(\vert {q}_\mathrm{r} \vert )\tilde{T}_{E}-\text{diag}(\tilde{\mathcal{T}}_{N1}\vert {q}_\mathrm{r}\vert )\tilde{T}_{N1}\label{subeq:nozerovol_nodes}\\
0 & = \tilde{\mathcal{T}}_{N0}^\top \text{diag}(\vert {q}_\mathrm{r} \vert )\tilde{T}_{E}-\text{diag}(\tilde{\mathcal{T}}_{N0}\vert {q}_\mathrm{r}\vert )\tilde{T}_{N0}, \label{subeq:zerovol_nodes}
\end{align}
\end{subequations}
where $\tilde{T}_E=\text{col}(\tilde{T}_{E,i})_{i\in \tilde{\mathcal{E}}}$ and $\tilde{V}_E=\text{col}(\tilde{V}_{E,i})_{i\in \tilde{\mathcal{E}}}$, and we recall that $q_\mathrm{r}=\text{col}(\tilde{q}_{\mathrm{E},i})_{i\in\tilde{\mathcal{E}}}$.

We note that the left-hand side of \eqref{subeq:zerovol_nodes} is zero since  $\tilde{V}_{N0}$ is the zero vector. Also, observe that \eqref{eq:redu_tem_dyn_comps} is a semi-explicit DAE. Moreover, since $\sum_{k  \in \tilde{\mathfrak{T}}_j} \vert \tilde{q}_{{\mathrm E},k}\vert >0$ for each $j\in \tilde{\mathcal{N}}_0$, then $\text{diag}(\tilde{\mathcal{T}}_{N0}\vert {q}_\mathrm{r}\vert )$ is non-singular. Then, \eqref{eq:redu_tem_dyn_comps} is a DAE with differentiation index one. This DAE can be turned into an ODE straightforwardly by clearing $\tilde{T}_{N0}$ from \eqref{subeq:zerovol_nodes} and substituting the result into \eqref{subeq:edges}. The resulting ODE is equivalent to \eqref{eq:TDYN_redu}, with $A_\mathrm{th}$ representing the Schur complement of the block $-\text{diag}(\tilde{\mathcal{T}}_{N0}\vert {q}_\mathrm{r}\vert )$ of
{\small
\begin{align*}
\tilde{\mathcal{A}}(\tilde{q}_\mathrm{r})=\begin{bmatrix}
-\text{diag}(\vert {q}_\mathrm{r}\vert) & \text{diag}(\vert {q}_\mathrm{r} \vert )\tilde{\mathcal{S}}_{N1}^\top & \text{diag}(\vert {q}_\mathrm{r} \vert )\tilde{\mathcal{S}}_{N0}^\top\\
\tilde{\mathcal{T}}_{N1}^\top \text{diag}(\vert {q}_\mathrm{r} \vert ) & -\text{diag}(\tilde{\mathcal{T}}_{N1}\vert {q}_\mathrm{r}\vert ) & 0\\
\tilde{\mathcal{T}}_{N0}^\top \text{diag}(\vert {q}_\mathrm{r} \vert ) & 0 & -\text{diag}(\tilde{\mathcal{T}}_{N0}\vert {q}_\mathrm{r}\vert )
\end{bmatrix}.
\end{align*}
}
The fact that $B_{\mathrm{pr}}P_\mathrm{pr}$ and $B_\mathrm{c}P_\mathrm{c}$ are obtained from the non-zero components of $\mathbf{P}_\mathrm{pr}$ and $\mathbf{P}_\mathrm{c}$ is direct.~$\hfill\blacksquare$
\end{pf}

\section{An extension to Corollary~1}\label{app:extension_corollary_1}

Corollary~\ref{corollary: thermal_state_dyn} can  be extended as follows. Assume that at any given time $t_z$,  $\sum_{k  \in \tilde{\mathfrak{T}}_j} \vert \tilde{q}_{{\mathrm E},k}(t_z)\vert =0$ for each node $j$ in some set $\Xi_{t_z}\subset \tilde{\mathcal{N}}_0$. Then,   $\tilde{T}_{N0,j}=\alpha$, for each  $j\in \Xi_{t_z}$,  and $\tilde{T}_{N0,k}=(\sum_{i\in \tilde{\mathfrak{T}}_k}\vert \tilde{q}_{\mathrm{E},i}\vert \tilde{T}_{\mathrm{E},i})/(\sum_{i\in \tilde{\mathfrak{S}}_k}\vert \tilde{q}_{\mathrm{E},i}\vert)$, for each $k\in \tilde{\mathcal{N}}_0\setminus \Xi_{t_z}$, represents an explicit solution to the algebraic constraints of the DAE \eqref{eq:redu_tem_dyn_comps}. Here $\alpha$ is an arbitrary real number.  

To see the above, let us write the $m$th component of \eqref{subeq:zerovol_nodes} as follows:
\begin{align}\label{eq:sol_fam_t_n0}
0=\sum_{i\in \tilde{\mathfrak{T}}_m}\vert \tilde{q}_{\mathrm{E},i}\vert \tilde{T}_{\mathrm{E},i}-\left(\sum_{i\in \tilde{\mathfrak{T}}_m}\vert \tilde{q}_{\mathrm{E},i}\vert \right)\tilde{T}_{\mathrm{N},m}.
\end{align}
On the one hand, we note that for  any $m\in \Xi_{t_z}$,  $\sum_{k  \in \tilde{\mathfrak{T}}_j} \vert \tilde{q}_{{\mathrm E},k}(t_z)\vert =0$, which implies $\tilde{q}_{\mathrm{E},i}=0$ for all $i\in \tilde{\mathfrak{T}}_m$. Then, $\tilde{T}_{\mathrm{N},m}=\alpha$, for any $\alpha\in \mathbb{R}$, makes \eqref{eq:sol_fam_t_n0} hold.  However, this choice has no effect  in  \eqref{subeq:edges}. Indeed,  due to mass balance at $m\in \Xi_{t_z}$, which reads as
\begin{align*}
\sum_{k  \in \tilde{\mathfrak{T}}_m} \vert \tilde{q}_{{\mathrm E},k}(t_z)\vert -\sum_{k  \in \tilde{\mathfrak{S}}_m} \vert \tilde{q}_{{\mathrm E},k}(t_z)\vert =0,~~~m\in \Xi_{t_z},
\end{align*}
the condition  $\tilde{q}_{\mathrm{E},i}=0$  for all $i\in \tilde{\mathfrak{T}}_m$, for all $m\in \Xi_{t_z}$,  means that the flow through any edge incident to $m\in \Xi_{t_z}$ is zero. Now, the  $n$th component of the term $\text{diag}(\vert \tilde{q}_E \vert )\tilde{\mathcal{S}}_{N0}^\top \tilde{T}_{N0}$ in \eqref{subeq:edges} is   given by
{\small
\begin{align*}
\left(\text{diag}(\vert \tilde{q}_E \vert )\tilde{\mathcal{S}}_{N0}^\top \tilde{T}_{N0}\right)_n & = \vert \tilde{q}_{\mathrm{E},n}\vert \left(\sum_{m \in \Xi_{t_z}} (\tilde{\mathcal{S}}_{N0})_{m,n}\tilde{T}_{\mathrm{N},m} \right.\\
& ~~~~ \left. +\sum_{m \in \tilde{\mathcal{N}}_0\setminus \Xi_{t_z}} (\tilde{\mathcal{S}}_{N0})_{m,n}\tilde{T}_{\mathrm{N},m}\right).
\end{align*}}
Then, we note  that $\tilde{T}_{\mathrm{N},m}$, with $m\in \Xi_{t_z}$, can only appear in the components of \eqref{subeq:edges} associated with the edges $n\in \tilde{\mathcal{E}}$ that  are incident to $m\in \Xi_{t_z}$. However, since these flows are zero, then \eqref{subeq:edges} is independent of $T_{\mathrm{N},m}$ for each $m\in \Xi_{t_z}$.

We conclude our argument by observing that for any $m\in \tilde{\mathcal{N}}\setminus \Xi_{t_z}$, it holds that  $\sum_{i\in \tilde{\mathfrak{T}}_m}\vert \tilde{q}_{\mathrm{E},i}\vert>0$. Then,  $\tilde{T}_{\mathrm{N},m}$ can be cleared from \eqref{eq:sol_fam_t_n0} and be substituted in \eqref{subeq:edges}, turning \eqref{subeq:edges} and \eqref{subeq:nozerovol_nodes} into an ODE.

\section{Further details of the proof of Lemma~\ref{lem:calA_neg_sem}}\label{app:further_details_proof_lemma_calAnegdef}

Here we treat the case in which some elements of the vector $\bar{q}_\mathrm{r}$ might be zero. We will use a notation (and procedure) analogous to what is used in Appendix~\ref{app:diff_index_1_noncontractedgraph}. Let $\tilde{\mathcal{E}}_0\subset \tilde{\mathcal{E}}$ the set of edges whose equilibrium flows are zero. Analogously, let $\tilde{\mathcal{E}}_1=\tilde{\mathcal{E}}\setminus \tilde{\mathcal{E}}_0$ denote the set of edges whose equilibrium flows are different from zero. Also, let $\tilde{\mathcal{N}}_0\subset \tilde{\mathcal{N}}$ the set of nodes which are not receiving any stream of water, {\em i.e.}, those nodes $k$ for which $\sum_{i\in \tilde{\mathfrak{T}}_k}\vert \bar{q}_{\mathrm{E},i}\vert =0$. Let $\tilde{\mathcal{N}}_1=\tilde{\mathcal{N}}\setminus \tilde{\mathcal{N}}_0$. Reorder $\tilde{\mathcal{E}}$ and $\tilde{\mathcal{N}}$ as $\tilde{\mathcal{E}}=\tilde{\mathcal{E}}_1\cup \tilde{\mathcal{E}}_0$ and $\tilde{\mathcal{N}}=\tilde{\mathcal{N}}_1\cup\tilde{\mathcal{N}}_0$. Extend such  an ordering (and labeling)   to any variable, parameter or matrix  associated with these sets ({\em e.g.,} $\tilde{T}_{\mathrm{E}}$ and $\tilde{V}_{\mathrm{E}}$, and $\tilde{\mathcal{S}}$ and $\tilde{\mathcal{T}}$).  In these circumstances, the matrix $\tilde{\mathcal{A}}(q_\mathrm{r})$ (see \eqref{eq:calA}) evaluated at $\bar{q}_\mathrm{r}$ can be written as a four-by-four  block matrix as shown in equation \eqref{eq:big_matrix}, where we have used the fact that  $\bar{q}_{\mathrm{r}0}$ and  $\tilde{\mathcal{T}}_{N0E1}\vert {q}_{\mathrm{r}1}\vert + \tilde{\mathcal{T}}_{N0E0}\vert {q}_{\mathrm{r}0}\vert $ are zero vectors. At this point we invoke \cite[Lemma~8]{hangos_thermo_99} to conclude that $\tilde{\mathcal{A}}(\bar{q}_\mathrm{r})$ is negative semi definite; note that the sub matrix formed by the non-zero blocks of $\tilde{\mathcal{A}}(\bar{q}_\mathrm{r})$ is a Kirchhoff's Convection Matrix when we consider that the components of  $\bar{q}_{\mathrm{r}1}$ and $\tilde{\mathcal{T}}_{N1E1} \vert \bar{q}_{\mathrm{r}1}\vert$ are non zero and we thus return to the case already treated in our proof to Lemma~\ref{lem:calA_neg_sem}.

\begin{figure*}[h]
\begin{align}\label{eq:big_matrix}
\mathcal{A}(\bar{q}_\mathrm{r}) & =\begin{bmatrix}
-\text{diag}(\vert \bar{q}_\mathrm{r1}\vert) & 0 & \text{diag}(\vert \bar{q}_\mathrm{r1} \vert )\tilde{\mathcal{S}}_{N1E1}^\top & \text{diag}(\vert \bar{q}_\mathrm{r1} \vert )\tilde{\mathcal{S}}_{N0E1}^\top\\
0 & -\text{diag}(\vert \bar{q}_\mathrm{r0}\vert)  & \text{diag}(\vert \bar{q}_\mathrm{r0} \vert )\tilde{\mathcal{S}}_{N1E0}^\top & \text{diag}(\vert \bar{q}_\mathrm{r0} \vert )\tilde{\mathcal{S}}_{N0E0}^\top\\
\tilde{\mathcal{T}}_{N1E1}^\top \text{diag}(\vert \bar{q}_\mathrm{r1} \vert ) & \tilde{\mathcal{T}}_{N1E0}^\top \text{diag}(\vert \bar{q}_\mathrm{r0} \vert )  & -\text{diag}(\tilde{\mathcal{T}}_{N1E1}\vert \bar{q}_{\mathrm{r}1}\vert + \tilde{\mathcal{T}}_{N1E0}\vert \bar{q}_{\mathrm{r}0}\vert ) & 0\\
\tilde{\mathcal{T}}_{N0E1}^\top \text{diag}(\vert \bar{q}_{\mathrm{r}1} \vert ) & \tilde{\mathcal{T}}_{N0E0}^\top \text{diag}(\vert \bar{q}_{\mathrm{r}0} \vert ) & 0 & -\text{diag}(\tilde{\mathcal{T}}_{N0E1}\vert \bar{q}_{\mathrm{r}1}\vert + \tilde{\mathcal{T}}_{N0E0}\vert \bar{q}_{\mathrm{r}0}\vert )
\end{bmatrix}\nonumber \\
& = \begin{bmatrix}
-\text{diag}(\vert \bar{q}_\mathrm{r1}\vert) & 0 & \text{diag}(\vert \bar{q}_\mathrm{r1} \vert )\tilde{\mathcal{S}}_{N1E1}^\top & 0\\
0 & 0 & 0 & 0\\
\tilde{\mathcal{T}}_{N1E1}^\top \text{diag}(\vert \bar{q}_\mathrm{r1} \vert ) & 0 & -\text{diag}(\tilde{\mathcal{T}}_{N1E1}\vert \bar{q}_{\mathrm{r}1}\vert ) & 0\\
0 & 0 & 0 &0
\end{bmatrix}
\end{align}
\end{figure*}

\section{On the usefulness of Theorems~\ref{theorem:shifted_passivity} and Theorem~\ref{theorem:shifted_pass_TDYN} for control design}\label{app:usefulness_passivity}

\subsection{Flow dynamics}
Based on the  results in \cite{bayu_scl_07,DePersis2011}, it is possible to claim that for the flow dynamics~\eqref{eq:flow_dynamics_qf},  a decentralized, proportional-integral controller of the form
\begin{subequations}\label{eq:pi_chords}
\begin{align}
w_{{\mathrm f},i}  & = -k^{\mathrm p}_{\mathrm{f},i} \left( q_{{\mathrm f},i}- q_{{\mathrm f},i}^\star \right) + z_{{\mathrm f},i} \\
\dot{z}_{{\mathrm f},i} & = -k^{\mathrm I}_{\mathrm{f},i}(q_{{\mathrm f},i}-  q_{{\mathrm f},i}^\star)
\end{align}
\end{subequations}
can achieve the global regulation of $q_{\mathrm f}$ towards any desired constant vector $ q_{\mathrm f}^\star$, subject to the assumption that each component of $w_{\mathrm f}$ is unconstrained and  $k^{\mathrm p}_{\mathrm{f},i},k^{\mathrm I}_{\mathrm{f},i}>0$ for all $i=1,\dots,{n_{\mathrm f}}$. A proof of an analogous claim can be found in \cite{jmnj_css_21}.

In some practical settings the flow  $q_{{\mathrm f},i}$ of the chord $i\in \mathcal{C}$ may not be  available for  measurement, but the  pressure drop, say $f_{{\mathrm f},i}(q_{{\mathrm f},i})$, of a valve adjacent to $i$ could be.  By considering  Lemma~\ref{lemma:-f_monotone},  we conjecture that the analysis procedure of \cite[Section~III]{DePersis2014}  can be adapted to our setting and  establish that a controller of the form
\begin{subequations}\label{eq:pi_chords_valves}
\begin{align}
w_{{\mathrm f},i}  & = -k^{\mathrm p}_{\mathrm{f},i}\left( \mu_{{\mathrm f},i} ( q_{{\mathrm f},i})-r_i \right) + z_{{\mathrm f},i} \\
\dot{z}_{{\mathrm f},i} & = -k^{\mathrm I}_{\mathrm{f},i} \left(\mu_{{\mathrm f},i}(q_{{\mathrm f},i})- r_i \right)
\end{align}
\end{subequations}
can achieve the global asymptotic regulation of the pressure drop $\mu_{{\mathrm f},i}(q_{{\mathrm f},i})$ towards any desired constant value $r_i$, for $i=1,\dots,n_{\mathrm f}$.

\subsection{Thermal dynamics}
{\color{black}By considering Theorem~\ref{theorem:shifted_pass_TDYN} and the results in  \cite{bayu_scl_07}, we claim that  it is possible to control the passive output of the thermal dynamics \eqref{eq:TDYN_redu} (assuming that $q_\mathrm{r}=\bar{q}_\mathrm{r}$) subject to the satisfaction of the  following detectability condition (see \cite{arjan_l2gain_book}):
\begin{equation}\label{eq:pi_thermal}
P_\mathrm{pr}=\bar{P}_\mathrm{pr},~y_{\mathrm{th}}= \bar y_{\mathrm{th}},~\forall t\geq 0~\Rightarrow \lim_{t\rightarrow \infty} T_\mathrm{th}=\bar T_\mathrm{th}.
\end{equation}
More precisely,  if \eqref{eq:pi_thermal} holds, then the  proportional-integral  controllers
\begin{subequations}\label{eq:pi_Tth}
\begin{align}
P_{{\mathrm{pr}},i}  & = -k^{\mathrm p}_{\mathrm{th},i} \left( y_{{\mathrm{th}},i}- y_{{\mathrm{th}},i}^\star \right) + z_{{\mathrm{th}},i} \\
\dot{z}_{{\mathrm{th}},i} & = -k^{\mathrm I}_{\mathrm{th},i} (y_{{\mathrm{th}},i}-  y_{{\mathrm{th}},i}^\star),
\end{align}
\end{subequations}
for $i=1,\dots,n_\mathrm{pr}$, achieve the asymptotic regulation of $y_{\mathrm{th}}$ towards any desired constant vector $ y_{\mathrm{th}}^\star$. Therefore, a classic  supply temperature control  scheme  can be enabled. 
}

\medskip

\bibliographystyle{abbrv}        

\balance

\bibliography{district_heating}           

\end{document}